\documentclass[12pt,preprint,useAMS,usenatbib]{emulateapj}
\usepackage{graphicx}
\usepackage{subfloat}
\usepackage{amsmath}
\def\OVI{\hbox{O~$\scriptstyle\rm VI$}}
\def\msun{\,{\rm M_\odot}}
\def\mdot{\,{\rm M_\odot yr^{-1}}}

\def\sfr{\,{\rm M_\odot\,yr^{-1}}}

\def\spose#1{\hbox to 0pt{#1\hss}}
\def\lta{\mathrel{\spose{\lower 3pt\hbox{$\mathchar"218$}}
     \raise 2.0pt\hbox{$\mathchar"13C$}}}
\def\gta{\mathrel{\spose{\lower 3pt\hbox{$\mathchar"218$}}
     \raise 2.0pt\hbox{$\mathchar"13E$}}}

\newcommand{\etal}{{et al.\ }}
\def\HI{\hbox{H~$\scriptstyle\rm I$}}
\def\iHI{\hbox{$H~\scriptstyle I$}}
\def\HII{\hbox{H~$\scriptstyle\rm II$}}

\def\CIV{\hbox{C~$\scriptstyle\rm IV$}}

\def\CII{\hbox{C~$\scriptstyle\rm II$}}

\def\OVI{\hbox{O~$\scriptstyle\rm VI$}}

\def\SiII{\hbox{Si~$\scriptstyle\rm II$}}

\def\SiIV{\hbox{Si~$\scriptstyle\rm IV$}}
\def\MgII{\hbox{Mg~$\scriptstyle\rm II$}}
\def\kms{\,{\rm km\,s^{-1}}}

\def\cmm{\,{\rm cm^{-2}}}
\def\cm3{\,{\rm cm^{-3}}}

\def\sfr{\,{\rm M_\odot\,yr^{-1}}}

\def\Lya{Ly$\alpha$}

\lefthead{Shen et al. 2013}
\righthead{Circumgalactic medium, galactic outflows, and cold streams}
\submitted{Submitted to the Astrophysical Journal}


\begin{document}

\title{The circumgalactic medium of massive galaxies at $z\sim 3$: a test for stellar feedback, galactic outflows, and cold streams}
\author{Sijing~Shen$^1$, Piero~Madau$^{1}$, Javiera~Guedes$^{2}$, Lucio Mayer$^3$, J. Xavier Prochaska$^{1,4}$, \& James Wadsley$^5$}

\altaffiltext{1}{Department of Astronomy and Astrophysics, University of California, 1156 High Street, Santa Cruz, CA 95064.}
\altaffiltext{2}{Institute for Astronomy, ETH Zurich, Wolgang-Pauli-Strasse 27, 8093 Zurich, Switzerland.}
\altaffiltext{3}{Institute of Theoretical Physics, University of Zurich, Winterthurerstrasse 190, CH-9057 Zurich, Switzerland.}
\altaffiltext{4}{UCO/Lick Observatory, University of California, 1156 High Street, Santa Cruz, CA 95064.}
\altaffiltext{5}{Department of Physics \& Astronomy, McMaster University, Hamilton, ON L8S 4M1, Canada.}

\begin{abstract}
We present new results on the kinematics, thermal and ionization state, and spatial distribution of metal-enriched gas in the circumgalactic medium (CGM)
of massive galaxies at redshift $\sim 3$, using the ``Eris" suite of cosmological hydrodynamic ``zoom-in" simulations. The reference run adopts a blastwave 
scheme for supernova feedback that produces large-scale galactic outflows, a star formation recipe based on a high gas density threshold, metal-dependent 
radiative cooling, and a model for the diffusion of metals and thermal energy. The effect of the local UV radiation field is added in post-processing. 
The CGM (defined as all gas at $R>0.2R_{\rm vir}=10$ kpc, where $R_{\rm vir}$ is the virial radius) contains multiple phases having a wide range of physical 
conditions, with more than half of its heavy elements locked in a warm-hot component at $T>10^5\,$K. Synthetic spectra, generated by drawing sightlines through 
the CGM, produce interstellar absorption line strengths of \Lya, \CII, \CIV, \SiII, and \SiIV\ as a function of galactocentric impact parameter 
(scaled to the virial radius) that are in broad 
agreement with those observed at high-redshift by \citet{Steidel10}. The covering factor of absorbing material declines less rapidly with 
impact parameter for \Lya\ and \CIV\ compared to \CII, \SiIV, and \SiII, with \Lya\ remaining strong ($W_{\rm Ly\alpha}>300\,$m\AA) to $\gta 5R_{\rm vir}=250$ 
kpc. Only about one third of all the gas within $R_{\rm vir}$ is outflowing. The fraction of sightlines within one virial radius that intercept optically 
thick, $N_{\rm HI}>10^{17.2}\,\cmm$ material is 27\%, in agreement with recent observations by \citet{Rudie12}. Such optically thick absorption is shown to trace
inflowing ``cold" streams that penetrate deep inside the virial radius. The streams, enriched to metallicities above 0.01 solar by previous episodes of 
star formation in the main host and in nearby dwarfs, give origin to strong ($N_{\rm CII}>10^{13}\cmm$) \CII\ absorption with a covering 
factor of 22\% within $R_{\rm vir}$ and 10\% within $2 R_{\rm vir}$. Galactic outflows do not cause any substantial suppression of the cold accretion mode. 
The central galaxy is surrounded by a large \OVI\ halo, with a typical column density $N_{\rm OVI}\gta 10^{14}\,\cmm$ and a near unity covering 
factor maintained all the way out to 150 kpc. This matches the trends recently observed in star-forming galaxies at low redshift by \citet{Tumlinson11}. 
Our zoom-in simulations of this single system appear then to reproduce quantitatively the complex baryonic processes that determine the exchange 
of matter, energy, and metals between galaxies and their surroundings. 
\end{abstract}


\section{Introduction} \label{intro}

Studies of the ionization, chemical, thermodynamic, and kinematic state of gaseous material in the circumgalactic medium (CGM) hold clues 
to understanding the exchange of mass, metals, and energy between galaxies and their surroundings. It is the poorly understood flows of 
gas into galaxies, and from them back into their environments, that determine the response of baryons to dark matter potential wells, 
regulate star formation, and enrich the intergalactic medium (IGM). Observations clearly show that galactic-scale metal-rich outflows with 
velocities of several hundred $\kms$ are ubiquitous in massive star-forming galaxies at high redshift \citep[e.g.,][]{Pettini01,Shapley03,Veilleux05,Weiner09}.
Far-UV spectra of close angular pairs of $z\sim$ 2-3 ``Lyman Break Galaxies" (LBGs) have recently provided a detailed map of 
such metal-enriched gas as a function of galactocentric distance \citep{Steidel10}, and a benchmark for simulations of the galaxy-IGM ecosystem.
The observed line strengths of several ionic species convey information about the physical conditions, covering fraction, and velocity 
spread of the absorbing multi-phase CGM on 100 kpc scales \citep{Steidel10}. These, together with the increasing amount of data that has been accumulated recently 
on the galaxy environs at high and low redshifts \citep[e.g.][]{Prochaska09,Crighton11,Prochaska11,Tumlinson11,Rudie12}, must be reproduced by hydrodynamical 
simulations of galaxy formation that aim at following the transport of heavy elements from their production sites into the IGM
\citep[e.g.][]{Oppenheimer08,Wiersma09,Shen10,Cen11}. 

Recent theoretical work has found that the majority of baryons accreted by galaxies below a certain critical mass flow in ``cold" rather than 
being shocked to the virial temperature \citep*[e.g.][]{Birnboim03,Keres05,Dekel06,Ocvirk08,Dekel09,Keres09}. This cold accretion mode occurs mainly along 
dense streams that deliver fresh fuel for star formation to the disk \citep[e.g.][]{Brooks09}. Even when a shock is present, cold gas accretion 
can occur along filaments that penetrate deep inside the hot halo \citep*[e.g.][]{Dekel06,Dekel09,Agertz09}. Signatures of such streams may include 
\Lya\ blobs \citep{Goerdt10}, Lyman Limit Systems \citep{Fumagalli11}, or \MgII\ absorbers \citep{Kacprzak12}. Previous numerical studies on cold flows and the 
cycle of baryons in and out of galaxies have either focused on a statistical description of how galaxies get their gas and sacrificed 
numerical resolution on individual galaxy scales \citep[e.g.][]{Keres05,Ocvirk08,Dekel09,Voort11}, 
or used ``zoom-in" simulations that do not generate strong galactic winds and hence cannot address the complex interaction between inflowing
and outflowing gas in the multiphase CGM \citep[e.g.][]{Faucher11,Fumagalli11,Kimm11,Stewart11,Goerdt12}. 

\begin{figure*}
\centering
\includegraphics[width=0.95\textwidth]{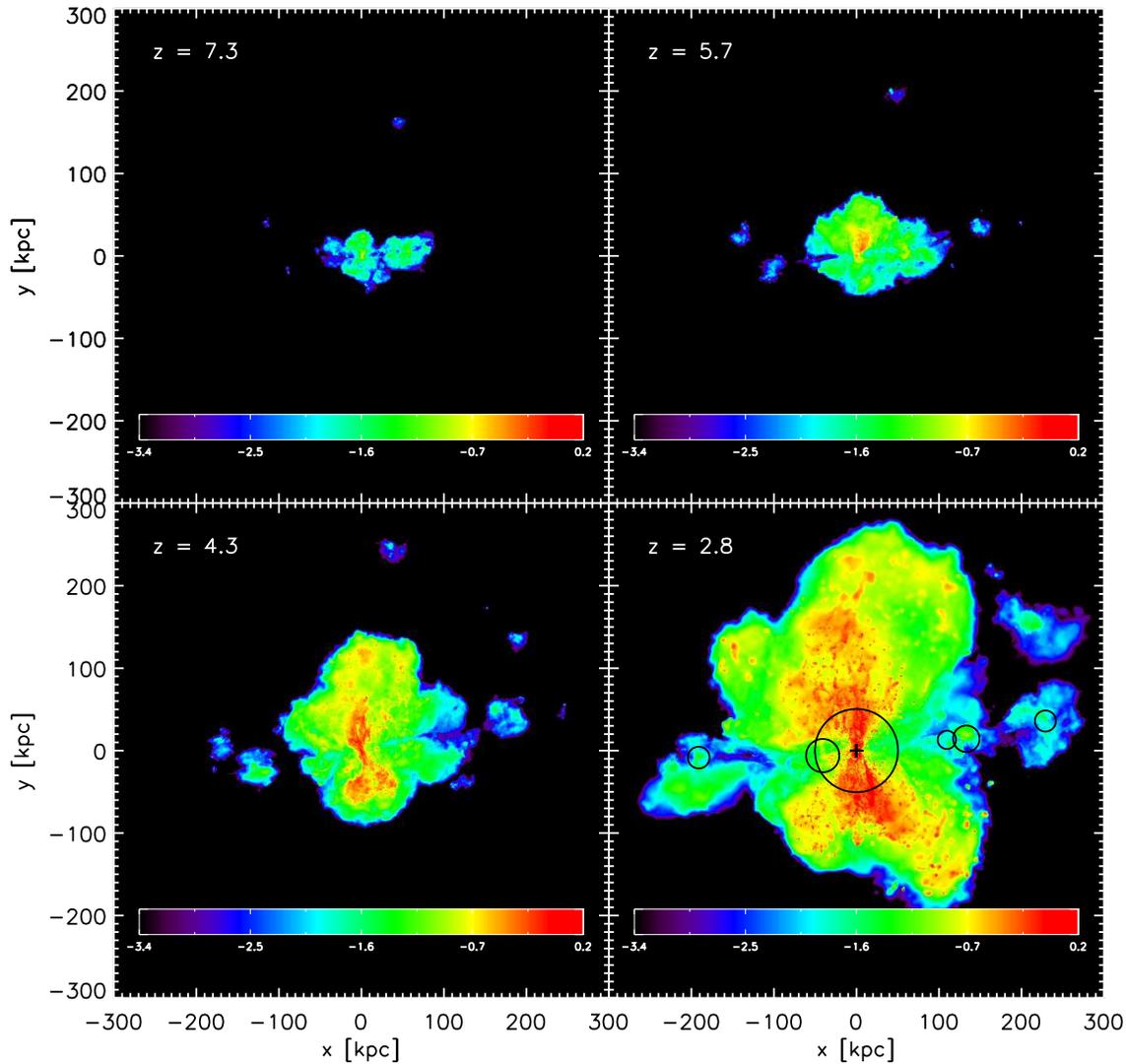}
\vspace{-0.3cm}
\caption{The growth of Eris2's metal enriched CGM. The figure shows the projected gas metallicity (i.e. the surface mass density of metals
divided by the total mass surface density) at different redshifts in 
a cube of 600 (proper) kpc on a side. The galaxy stellar disk is seen nearly edge-on in this projection. At $z=2.8$, the galaxy center 
is indicated by the plus sign at coordinates $(x,y) = (0,0)$, and its virial radius is marked by the black circle. Also marked are the 
virial radii of the 5 most massive nearby dwarfs and satellites. The metallicity is given in units of solar, 
$Z_\odot=0.0142$ \citep{Asplund09}.  
}
\label{fig1}
%\vspace{-0.1cm}
\end{figure*}

In \citet{Shen12}, we presented a high-resolution analysis of the metal-enriched CGM of a massive galaxy at $z=3$ using results from the ``Eris" 
suite of new cosmological hydrodynmic ``zoom-in" simulations, in which a close analog of a Milky Way system arises at the present epoch \citep{Guedes11}. 
Such simulations adopt a blastwave scheme for supernova feedback that -- in combination with a star formation recipe based on a high gas density 
threshold -- generates galactic outflows without explicit wind particles. \citet{Shen12} identified three sources of heavy elements within 
200 physical kpc from the galaxy's center: the main host, its satellite progenitors (accreted by the main host before redshift 3), 
and nearby dwarfs. The CGM was shown to be a blend of 
outflowing metal-rich and infalling metal-poor gas, with approximately half of all gas-phase metals locked in a hot-warm component. In this 
Paper we use a new simulation of the ``Eris" suite to create synthetic far-UV absorption spectra through the CGM and compare them to 
high-redshift observations, and to make predictions about the properties and detectability of cold accretion flows. To anticipate the results of our 
study, the simulated transmission spectra are found to be in good agreement with the interstellar metal absorption line strengths and \HI\ covering 
factors as a function of impact parameter recently measured in a $z=$2-3 galaxy sample by \citet{Steidel10} and \citet{Rudie12}. We show that cold 
accretion streams within 1-2 $R_{\rm vir}$ are traced by optically thick, metal-enriched absorbers with covering factor of 10-20\%.
There is no substantial suppression of the cold accretion mode caused by galactic outflows. The central galaxy is surrounded by a large halo of 
collisionally-ionized \OVI\ plasma that matches the properties of the \OVI\ bubbles recently detected around low-redshift starf-forming galaxies by \citet{Tumlinson11}.

\begin{figure}[h]
\centering
\includegraphics[width=0.49\textwidth]{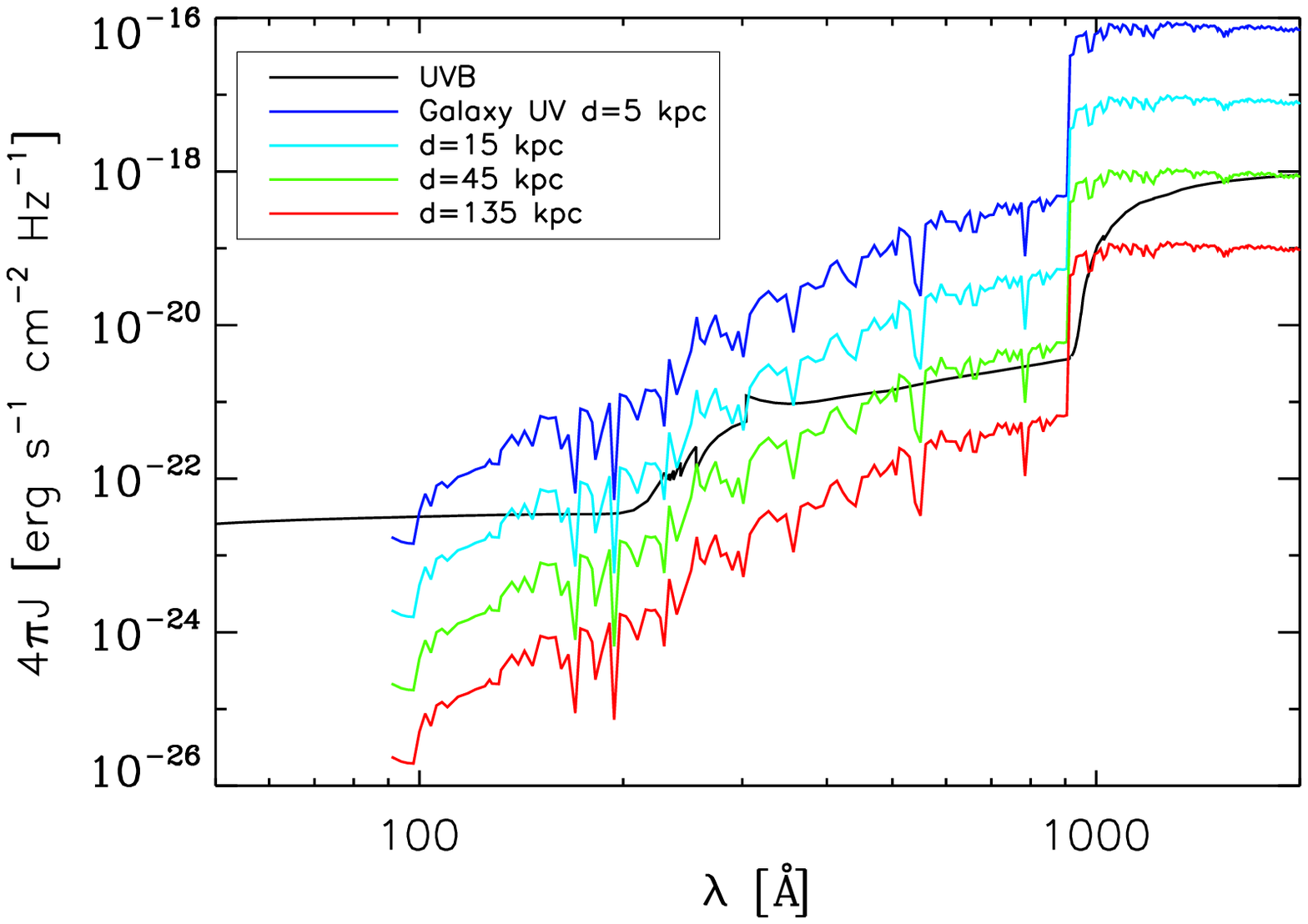}
\vspace{-0.2cm}
\caption{The far-UV flux impinging on Eris2's CGM. The solid line shows the $z=2.8$ UV diffuse background from \citet{Haardt12}, while the colored 
lines show the local stellar radiation at different distances from the center, assuming an escape fraction of $f_{\rm esc}=3\%$ for the ionizing 
photons that leak from the star-forming regions into the CGM. The synthetic galaxy spectrum was produced running \textsc{Starburst99} \citep{Leitherer99} 
assuming a (constant) star formation rate of $20\,\sfr$ and a \citet{Kroupa01} IMF. 
}
\label{fig2}
\vspace{0.2cm}
\end{figure}

\section{Simulation} 
\label{simulation}

The Eris suite of zoom-in simulations of a Milky Way galaxy analog is being performed in a {\it Wilkinson Microwave Anisotropy Probe} 3-year 
cosmology with the parallel TreeSPH code \textsc{Gasoline} \citep{Wadsley04}. Details of main run were given in \citet{Guedes11}. 
The high-resolution region, 1 Mpc on a side at $z=0$, contains 13 million dark matter particles and an equal number of gas particles, for a final 
dark and gas particle mass of $m_{\rm DM}=9.8\times 10^4\,\msun$ and $m_{\rm SPH}=2\times 10^4\,\msun$, respectively. The spline softening length is 120 pc.
The twin run analyzed here (``Eris2") presents several improvements over the previously published versions \citep{Guedes11,Shen12}:
1) the initial stellar mass function (IMF) follows the modern determination by \citet{Kroupa01}. This increases the number
of Type II supernovae per unit stellar mass by about a factor of 2, and the IMF-averaged metal yield by a factor of 3 compared to \citet{Kroupa93};  
2) the metallicity-dependent radiative cooling at all temperatures in the range 100-$10^9$ K is determined using pre-computed tabulated 
rates from the photoionization code \textsc{Cloudy} \citep{Ferland98}, following \citet{Shen10}. \textsc{Cloudy} tables assume that metals are in ionization equilibrium.
The ionization, cooling, and heating rates for primordial species (H, H$^+$, He, He$^+$, He$^{++}$) are calculated time-dependently from the rate equations.  
A uniform, cosmic UV background modifies the ionization and excitation state of the gas, photoionizing away abundant metal ions and reducing the cooling efficiency. 
It is implemented using the new \citet{Haardt12} redshift-dependent spectra, 
including emission from quasars and star-forming galaxies. A non-uniform local stellar radiation field is added in post-processing, as detailed below. As in previous runs, the 
gas is assumed to be optically thin to ionizing radiation at all wavelengths; 3) star formation occurs stocastically when cold ($T<10^4$ K), virialized gas reaches a threshold 
density of $n_{\rm SF}=20$ atoms cm$^{-3}$ (4 times higher than used in the original Eris run), and proceeds at a rate $d\rho_*/dt=0.1 \rho_{\rm gas}/t_{\rm dyn}$, where $\rho_*$ 
and $\rho_{\rm gas}$ are the stellar and gas densities, and $t_{\rm dyn}$ is the local dynamical time; 4) star particles inject energy, mass, and metals back into
the ISM through Type Ia and Type II SNe and stellar winds. We track the formation of Oxygen and Iron separately, and convert Oxygen to alpha-elements and Iron to iron-peak
elements assuming solar abundances patters \citep{Asplund09}. In the absence of some implementation of diffusion, metals are locked into specific particles and their
distribution is artificially inhomogeneous \citep{Wiersma09}. Eris2 includes a scheme for turbulent mixing in shearing flows that redistributes heavy elements and thermal energy 
between wind material and the ambient gas. Following \citet{Shen10}, the mixing of any scalar quantity $A$ is approximated by a diffusion term
\begin{equation}
\left({dA\over dt}\right)_D  =  \nabla (D \nabla A),
\end{equation}
where the diffusion coefficient $D$ follows the model proposed by \citet{Smagorinsky63} for the atmospheric boundary layer,
\begin{equation}
D = C\ |S_{ij}| \ h^2.
\end{equation}
Here $S_{ij}$ is the trace-free velocity shear tensor and $h$ is the measurement scale (taken to be equal to $h_{\rm SPH}$). This choice for $S_{ij}$
results in no diffusion for compressive or purely rotating flows. As in \citet{Shen10}, we choose a coefficient value of $C=0.05$ expected from turbulence theory.
This diffusion is applied to both thermal energy and heavy elements. Metal mixing was not implemented in the Eris and ErisMC simulations of \citet{Guedes11}
and \citet{Shen12}. The impact of metal diffusion is to spread heavy elements from outflowing wind material to the surroundings so that more gas 
gets contaminated at low levels \citep{Shen10}. 

As in previous simulations of this series, the feedback scheme follows the recipes of \citet{Stinson06}. 
Each Type II SN deposits metals and a net energy of $0.8 \times 10^{51}\,$ergs into a `blastwave radius', and 
the heated gas has its cooling shut off (to model the effect of feedback at unresolved scales) until the end of the momentum-conserving phase of the 
blastwave, which is set by the local gas density and temperature and by the total amount of energy injected \citep{McKee77}. No kinetic 
energy is explicitly assigned to particles within the radius of the blastwave. The energy injected 
by many SNe adds up to create larger hot bubbles and longer shutoff times. The main difference of this feedback model compared to 
other ``sub-grid" schemes \citep[e.g.][]{Springel03} is that it keeps galactic outflows hydrodynamically coupled to the energy injection by 
SNe. In combination with a high gas density threshold for star formation (which enables energy deposition
by SNe within small volumes), this scheme has been found to be key in producing realistic dwarf galaxies \citep{Governato10} and late-type massive spirals   
\citep{Guedes11}. {We stress that the parameters of the simulation have not been finely tuned to provide a fit to any of the observations discussed in this Paper. 
The modifications implemented in Eris2 compared to previously published runs of this series have been simply dictated by improving and updating the physics
of galaxy formation and the CGM.}  

\subsection{Local radiation field} 

Aside from the isotropic cosmological UV background, Eris2's CGM is photoionized by a local non-uniform stellar radiation field.
We have run \textsc{Starburst99} \citep{Leitherer99} and produced a synthetic galaxy stellar spectrum assuming star formation proceeds continuously
at a constant rate of $20\,\sfr$ and a \citet{Kroupa01} IMF. The galaxy spectrum, multiplied by the distance dilution factor $(4\pi d^2)^{-1}$ (for $d>0.9$ kpc, where $d$ is the
distance from the center of Eris2) and by the frequency-independent (absorption cross-section weighted) fraction of ionizing photons that leaks from the star-forming regions 
into the CGM and IGM, $\langle f_{\rm esc}\rangle$,     
has been added to the UV background to create a non-uniform radiation field. This approximation places all the stellar sources 
within the inner kpc, and breaks down for distances comparable to the scale length of the stellar disk. We have assumed an escape fration of 
$f_{\rm esc}=3\%$ (close to the recent upper limit, 5\%, for $z\approx 3$ LBGs derived by \citealt{Boutsia11}) and run the photoionization code 
\textsc{Cloudy} to calculate again, {\it in post-processing}, ionization fractions for an optically thin slab of gas at the density,
temperature, metallicity, and impinging radiation field of the simulated SPH particles, under the assumption that the temperature of the gas is not modified by the local 
radiation field.  

\begin{subfigures}

\begin{figure*}
\centering
\includegraphics[width=0.9\textwidth]{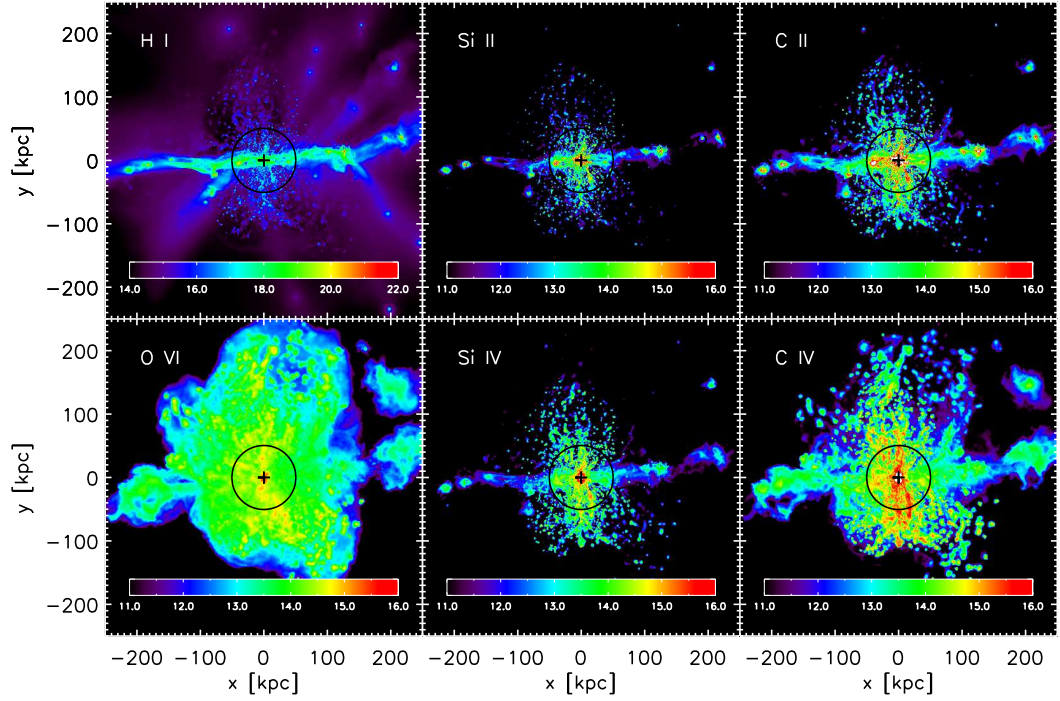}
\vspace{-0.3cm}
\caption{A map of the projected column density in a cube of 500 (proper) kpc on the side from the Eris2 simulation at $z=2.8$. 
The 6 panels show \HI, \CII, \CIV, \SiII, \SiIV, and \OVI. Intervals of column density in the range $10^{11}-10^{22}\,\cmm$ for \HI\ and 
$10^{11}-10^{16}\,\cmm$ for all metal ions are marked in the panels with different colors.
}
%\vspace{+0.2cm}
\label{fig3a}
\end{figure*}

\begin{figure*}
\centering
\includegraphics[width=0.9\textwidth]{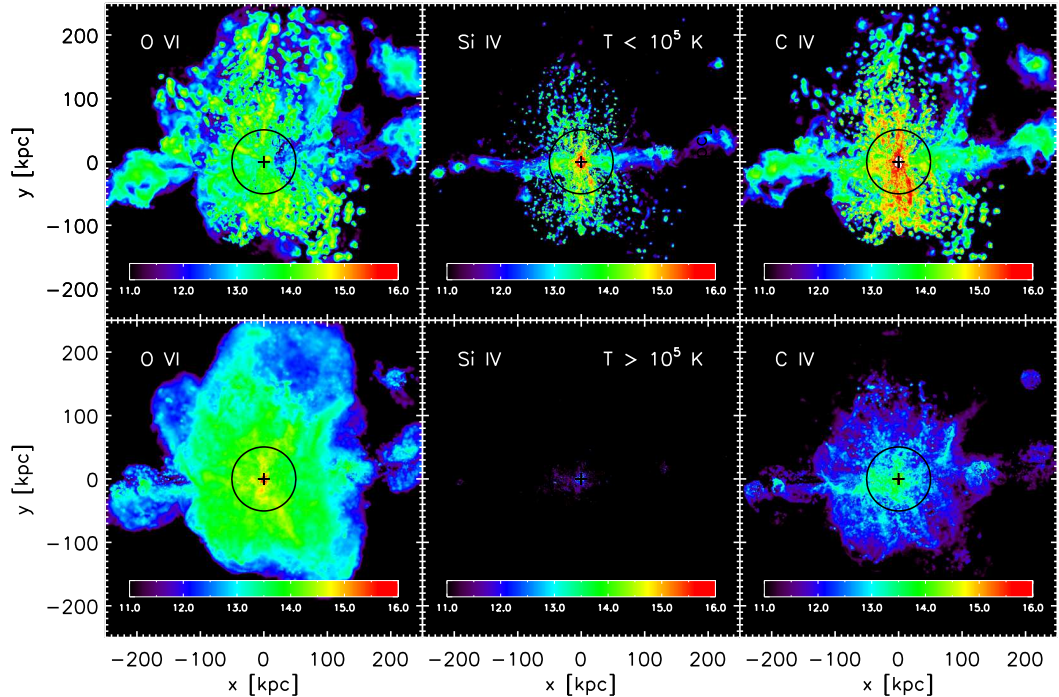}
\vspace{-0.3cm}
\caption{The multi-phase nature of Eris2's CGM. Same as Fig. \ref{fig3a} but for cool ($T<10^5$ K, {\it top panel}) and warm-hot ($T>10^5$ K, {\it bottom panel}) gas 
only, as traced by the high-ionization species \CIV, \SiIV, and \OVI. 
}
%\vspace{+0.2cm}
\label{fig3b}
\end{figure*}

\end{subfigures}

\begin{subfigures}
\begin{figure*}
\centering
\includegraphics[width=0.9\textwidth]{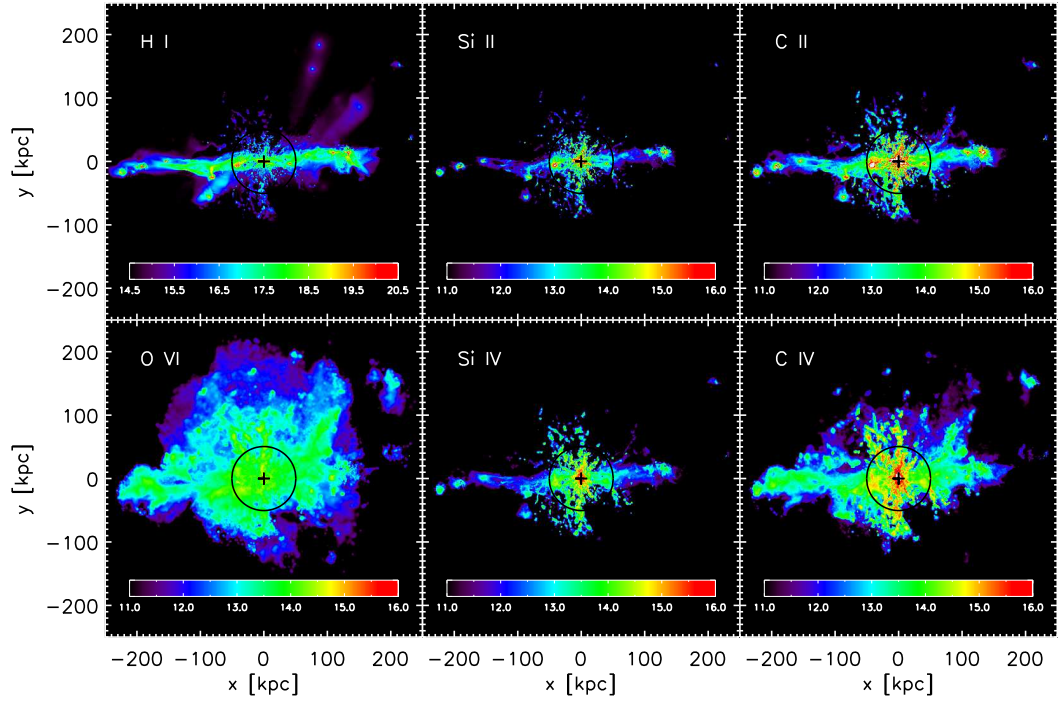}
\vspace{0.0cm}
\caption
{The kinematics of Eris2's CGM. Same as Fig. \ref{fig3a} but for inflowing gas only, as traced by low- and high-ionization species.
}
%\vspace{+0.2cm}
\label{fig4a}
\end{figure*}

\begin{figure*}
\centering
\includegraphics[width=0.9\textwidth]{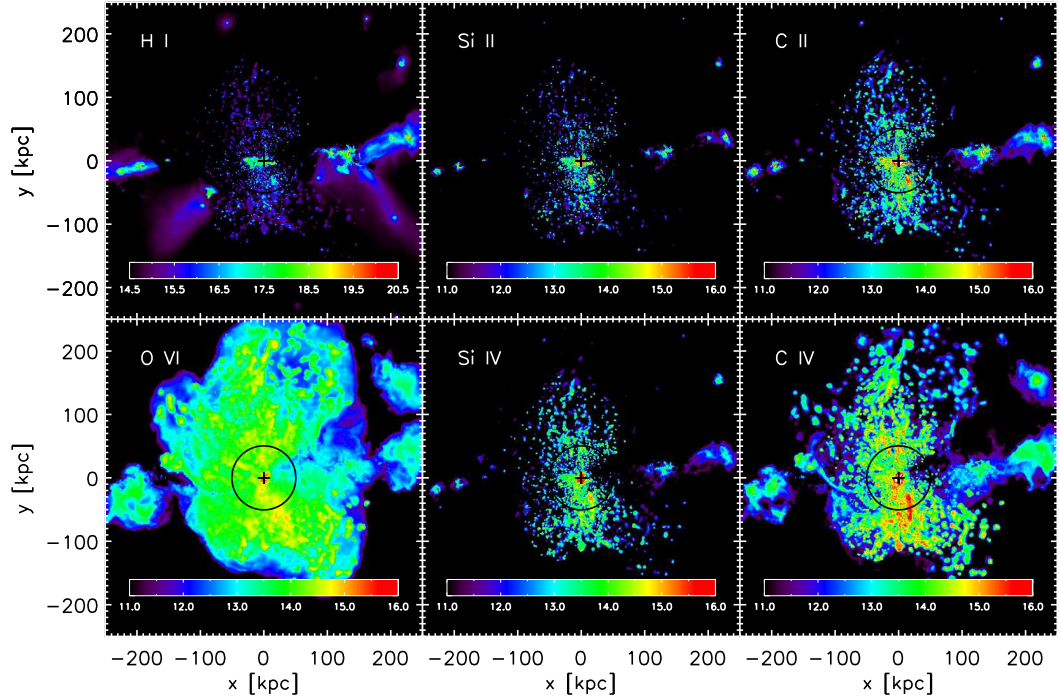}
\vspace{0.0cm}
\caption{Same as Fig. \ref{fig4a} but for outflowing gas only.
}
%\vspace{+0.2cm}
\label{fig4b}
\end{figure*}

\end{subfigures}

\begin{figure*}
\centering
\includegraphics[trim=0.3in 2.5in 0in 3in,clip,width=0.99\textwidth]{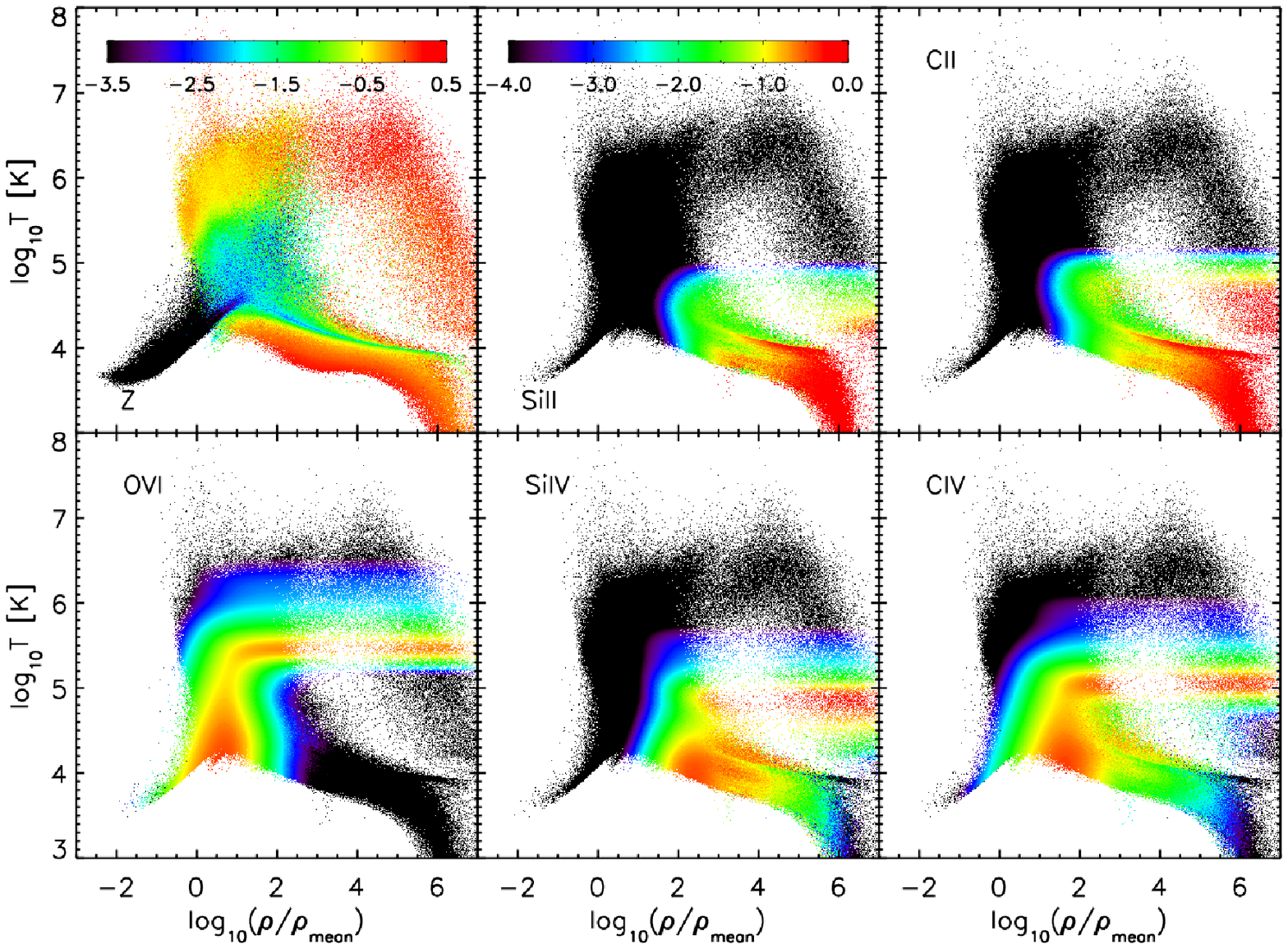}
\vspace{0.0cm}
\caption{Distribution of all enriched gas in the temperature-density plane at $z=2.8$ within the simulation volume. The top left panel shows the mass-weighted 
total metallicity (in units of solar), while in all the other panels the color coding indicates the ion fractions for the species \CII, \CIV, \SiII, \SiIV, and \OVI. 
The mean baryon density of the universe at this redshift is $1.4\times 10^{-5}\,\cm3$. Hot enriched gas vented out in the halo by the cumulative effect 
of SN explosions can be seen cooling (first adiabatically then radiatively) and raining back onto the disk in a ``galactic fountain". 
}
\label{fig5}
\vspace{+0.cm}
\end{figure*}

\section{Analysis and properties of the CGM}\label{CGM}

At $z=2.8$ (the lowest redshift reached by our simulation), Eris2 has a virial mass of $M_{\rm vir} = 2.6 \times 10^{11}\,\msun$, a virial radius of $R_{\rm vir} = 50$ kpc, 
a stellar mass of $M_*=1.5\times 10^{10}\,\msun$ (a factor 1.4 smaller than in ErisMC, see \citealt{Shen12}) and is forming stars at a rate of 
${\rm SFR}=20\,\sfr$. The mass loading factor at $R_{\rm vir}$ (characterizing the amount of material involved in the galactic outflow),
$\eta = \dot M_w/{\rm SFR}$ where $\dot M_w$ is the rate at which mass is ejected, ranges from 0.2 to 0.9 in the redshift interval 2.8-3.2. 
Observations of galactic outflows powered by starbursts suggest a wide range of mass loading factors, $\eta = 0.01 - 10$, with no obvious 
correlation with the star formation rates of their hosts \citep{Veilleux05}. While in low-resolution cosmological simulations, the mass loading 
factor is one of the input parameters \citep[e.g.,][]{Oppenheimer08}, in our SN-driven blastwave feedback scheme there is no specific 
parameter for mass loading.

Eris2's baryon fraction is 0.13, and the mean gas and 
stellar metallicities within $R_{\rm vir}$ are $\langle Z_g\rangle =0.7\,Z_\odot$ and $\langle Z_*\rangle =Z_\odot$, 
respectively. The galaxy's \HII\ regions (defined by all gas at $T=10^4$ K within 10 kpc from the center) are characterized by $12 +\log 
{\rm (O/H)}=8.5$, in agreement with the value, $12 +\log {\rm (O/H)}=8.42\pm 0.06$, 
measured by \citet{Erb06a} in star-forming galaxies of the same stellar mass at $\langle z\rangle=2.2$. 
About half of all the gas-phase heavy elements reside in Eris2's CGM (defined as all gas at $R>0.2R_{\rm vir}=10$ kpc), and
about half of all the CGM metals are locked in a warm-hot component at $T>10^5\,$K. 
The total mass of processed material with $Z>10^{-3}\,Z_\odot$ within 150 kpc is $4.7\times 10^{10}\,\msun$, while the 
mass of more pristine, $Z<10^{-3}\,Z_\odot$, gas within the same region is $1.04\times 10^{10}\,\msun$.
% NB If we use Z = 0 exactly, the mass is only 7.3e7 Msun, about 3500 particles. Metal diffusion contaminated a lot gas with very low metallicity. 
It is this region that will be compared in \S~\ref{comparison} to observations of the CGM around LBGs.
Figure \ref{fig1} shows the projected gas-phase metallicity in a cube 500 (physical) kpc on a side at different epochs. 
While metal-enriched material is seen as far as $250$ kpc from the center, within the same region we identify 11
dwarf satellite systems with masses above $10^9\,\msun$ that are also forming stars and polluting the CGM of Eris2 \citep{Shen12,Porciani05}. 
Figure \ref{fig2} shows the far-UV flux impinging on Eris2's CGM. With the adopted escape fraction, the galaxy flux 
at 1 Ryd dominates over the diffuse UV background at all distances $<45$ kpc. The most noticeable impact of the stellar radiation field will be to increase
(compared to a case with $f_{\rm esc}=0$) the interstellar absorption line strengths of \SiIV\ and \CIV\ and to decrease the equivalent
width of hydrogen \Lya\ at small impact parameters.

The processed simulation provides positions, velocities, abundances, and temperatures of the baryonic fluid elements represented by SPH particles.
To study the CGM of Eris2, we draw through the simulation volume $500\times 500$ regularly spaced sightlines within a projected sitance of 250 physical kpc 
from the center, for three orthogonal projections of the galaxy. Each particle is spread over its 3-dimensional SPH smoothing kernel, and a line integral is taken through 
the smoothed distribution to determine the line absorption optical depth as a function of frequency. For definiteness, the gas-phase column 
density, $N$, of an absorbing ion along a sightline of length $L$, is calculated as  
\begin{equation}
\begin{split}
N &= \displaystyle\sum\limits_j \int_0^L (m_j Z_j/m)W(r_{jl},h_j)dl\\
&= \displaystyle\sum\limits_{j} (m_j Z_j/m)W_{\rm 2D}(r_{jl},h_j), 
\end{split}
\label{eqcolumn}
\end{equation}
where $m_j$ is the gas particle mass, $Z_j$ its mass fraction in the relevant ion, and $m$ is the atomic mass of the ion.         
This amounts to a line integral through the 3D smoothing kernel $W(r_{jl},h_j)$ of each particle whose smoothing volume $V_j=(4/3)\pi h_j^3$
is pierced by the sightline at impact parameter $r_{jl}$. In the equation above we indicate with $W_{\rm 2D}$ the 2D smoothing kernel (with units of length$^{-2}$).

\begin{figure}
\centering
\includegraphics[width=0.49\textwidth]{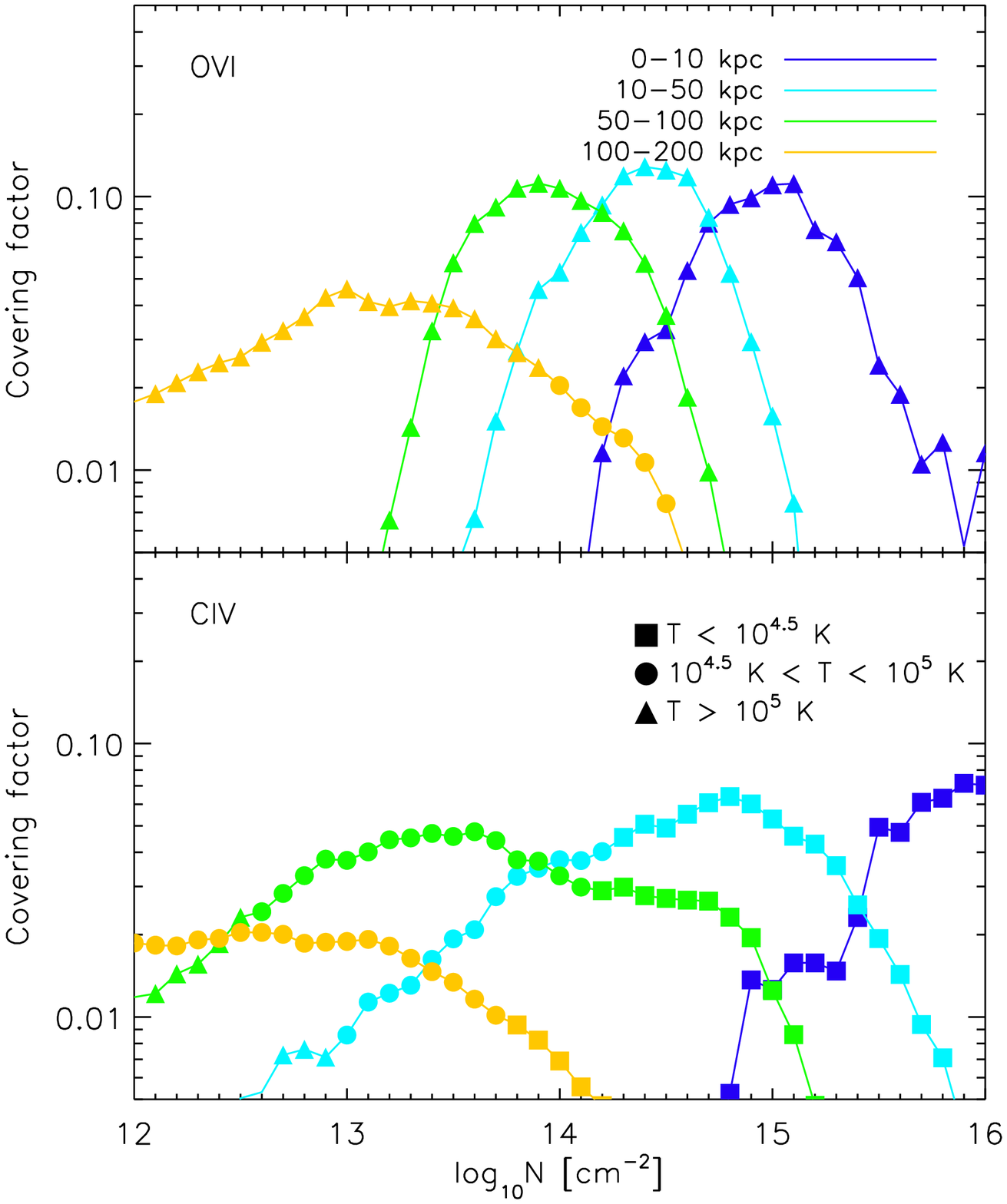}
\vspace{+0.2cm}
\caption{The covering factor (defined as the percentage of lines of sight with a given column density of absorbing gas, as seen by the observer) 
of \OVI\ ({\it top}) and \CIV\ ({\it bottom}) absorbing material through Eris2's CGM as a function of column density $N$. 
Values are plotted in bins of $\Delta\log N=0.1$. 
The color coding marks covering factor in different ranges of impact parameter, 0-10 kpc ({\it dark blue}), 10-50 kpc ({\it cyan}), 
50-100 kpc ({\it green}), 100-200 kpc ({\it yellow}). Different symbols are used to indicate the median temperature of a given column of 
absorbing gas over all sightlines: $T<10^{4.5}$ K ({\it squares}), $10^{4.5}<T<10^5$ K ({\it circles}), and $T>10^5$ K ({\it triangles}).    
}
\label{fig6}
\end{figure}

Because of the assumed spline smoothing \citep{Monaghan92}, only SPH particles at impact parameter $r_{jl}<2h_j$ contribute to the line integral
in equation (\ref{eqcolumn}). Each such particle is characterized by a velocity $v_j$ along the sightline, and by a temperature $T_j$, so that
the total optical depth at the observed frequency $\nu$ in a given line transition can be written as
\begin{equation}
	\tau(\nu) = \displaystyle\sum\limits_{j} (m_j Z_j/m)W_{\rm 2D}(r_{jl},h_j)\sigma_{j}(\nu), 
\label{tau_dis}
\end{equation} 
and the integrated absorption equivalent width of the line is 
\begin{equation}
	W_0= {c\over \nu_0^2} \int [1-e^{-\tau(\nu)}]d\nu.
\label{EW}
\end{equation}  
Here, the absorption cross-section is expressed in terms of the Voigt function as follows:
\begin{equation}
	\sigma_{j}(\nu) =\frac{sa_j}{{\pi}^{3/2} \Delta \nu_{j}} \int_{-\infty}^{+\infty}dy \frac{\exp(-y^{2})}{(x_j-y)^2 + a_j^2}, 
\label{sigma}
\end{equation}  
where $s$ is the frequency integrated absorption cross-section (proportional to the oscillator strength $f$ of the transition), $\Delta \nu_{j} 
\equiv \nu_0(b_j/c)$ is the Doppler width, $a_j \equiv \Gamma/(4\pi\Delta \nu_{j})$, $x_j \equiv (\nu-\nu_j)/(\Delta \nu_{j})$, $\nu_j\equiv \nu_0(1-v_j/c)$ 
is the line center frequency corrected for the gas velocity along the line of sight, $b_j  \equiv \sqrt{2kT_j/m}$ is the Doppler parameter,
and all other symbols have their usual meaning. We assume that the galaxy is at zero velocity, and include the contribution of bulk motion 
and Hubble expansion to the particle velocity $v_j$. The flux decrement in equation (\ref{EW}) is integrated over $\pm 500\,\kms$ centered on the galaxy's position, 
and each line of sight passes through a simulation volume 1 Mpc (physical) across.     

\begin{subfigures}

\begin{figure*}
\centering
\includegraphics[width=0.8\textwidth]{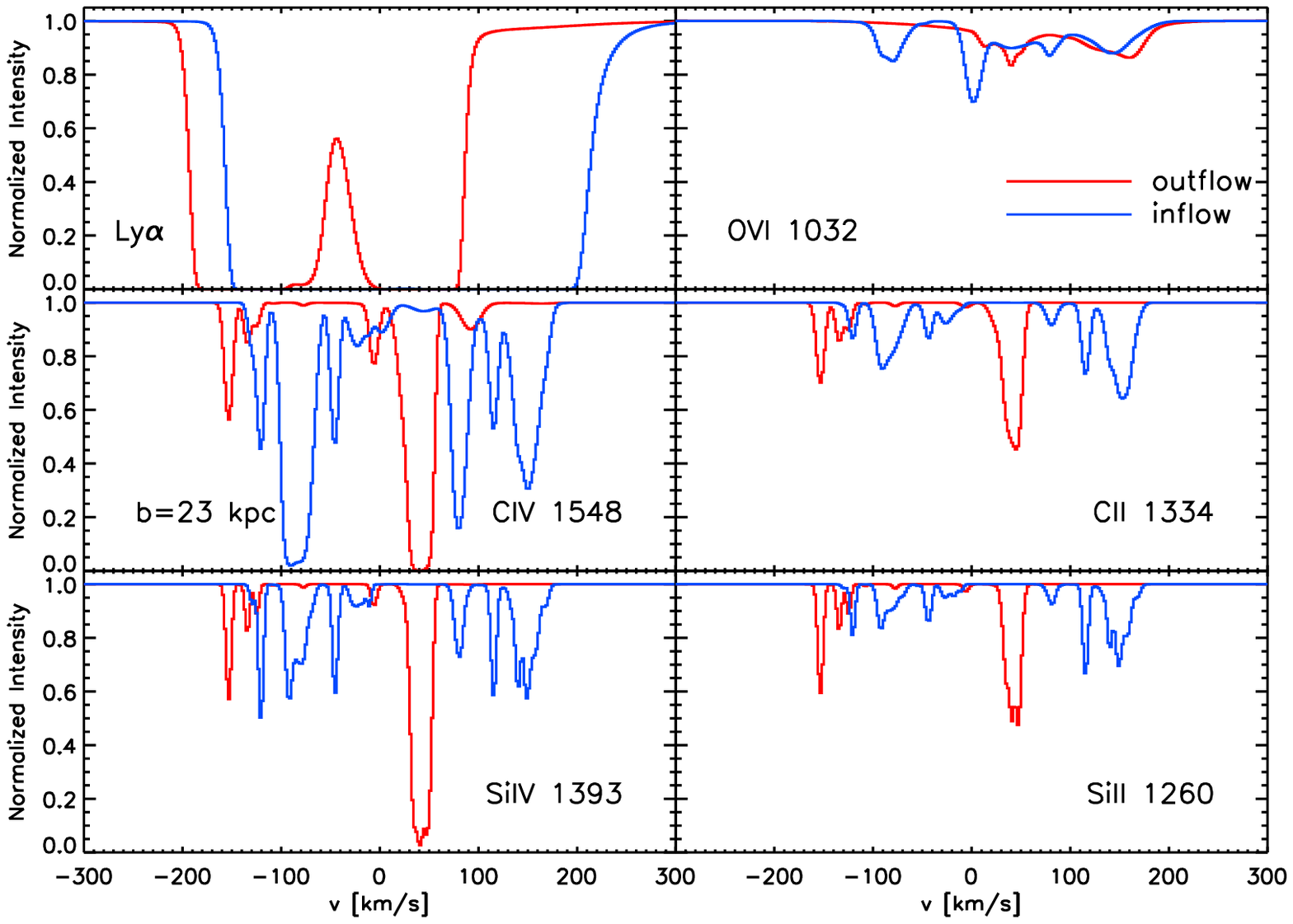}
\vspace{+0.2cm}
\caption{A simulated absorption spectrum through Eris2's CGM at $z=2.8$, computed at a resolution of $2\,\kms$ and plotted in velocity space. 
A random sightline at impact parameters $b=0.46 R_{\rm vir}=23$ kpc from the galaxy center is shown. Gas with positive radial velocity with respect to systemic is redshifted.
The total column densities of the absorbing \HI, \CII, \CIV, \SiII, \SiIV, and \OVI\ ions are, respectively $2.6\times 10^{16}, 1.1\times 10^{14},
4.2\times 10^{14}, 8.7\times 10^{12}, 5.3\times 10^{13},$ and $1.3\times 10^{14}\,\cmm$.   
}
\label{fig7a}
\end{figure*}

\begin{figure*}
\centering
\includegraphics[width=0.8\textwidth]{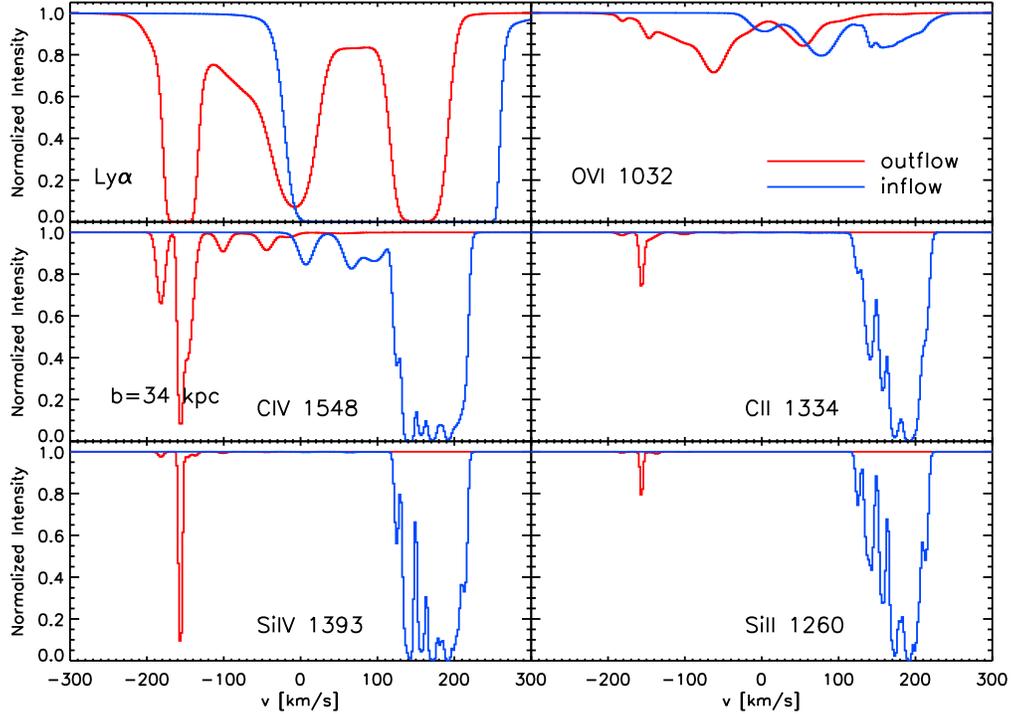}
\vspace{+0.2cm}
\caption{Same as Fig. \ref{fig7a}, but for a sightline at impact parameter $b=34\,$kpc.
The total column densities of the absorbing \HI, \CII, \CIV, \SiII, \SiIV, and \OVI\ ions are, respectively $1.3\times 10^{17}, 4.2\times 10^{14},
4.6\times 10^{14}, 3.7\times 10^{13}, 1.3\times 10^{14},$ and $1.8\times 10^{14}\,\cmm$.    
}
\label{fig7b}
\end{figure*}

\end{subfigures}

\begin{figure}
\centering
\includegraphics[width=0.49\textwidth]{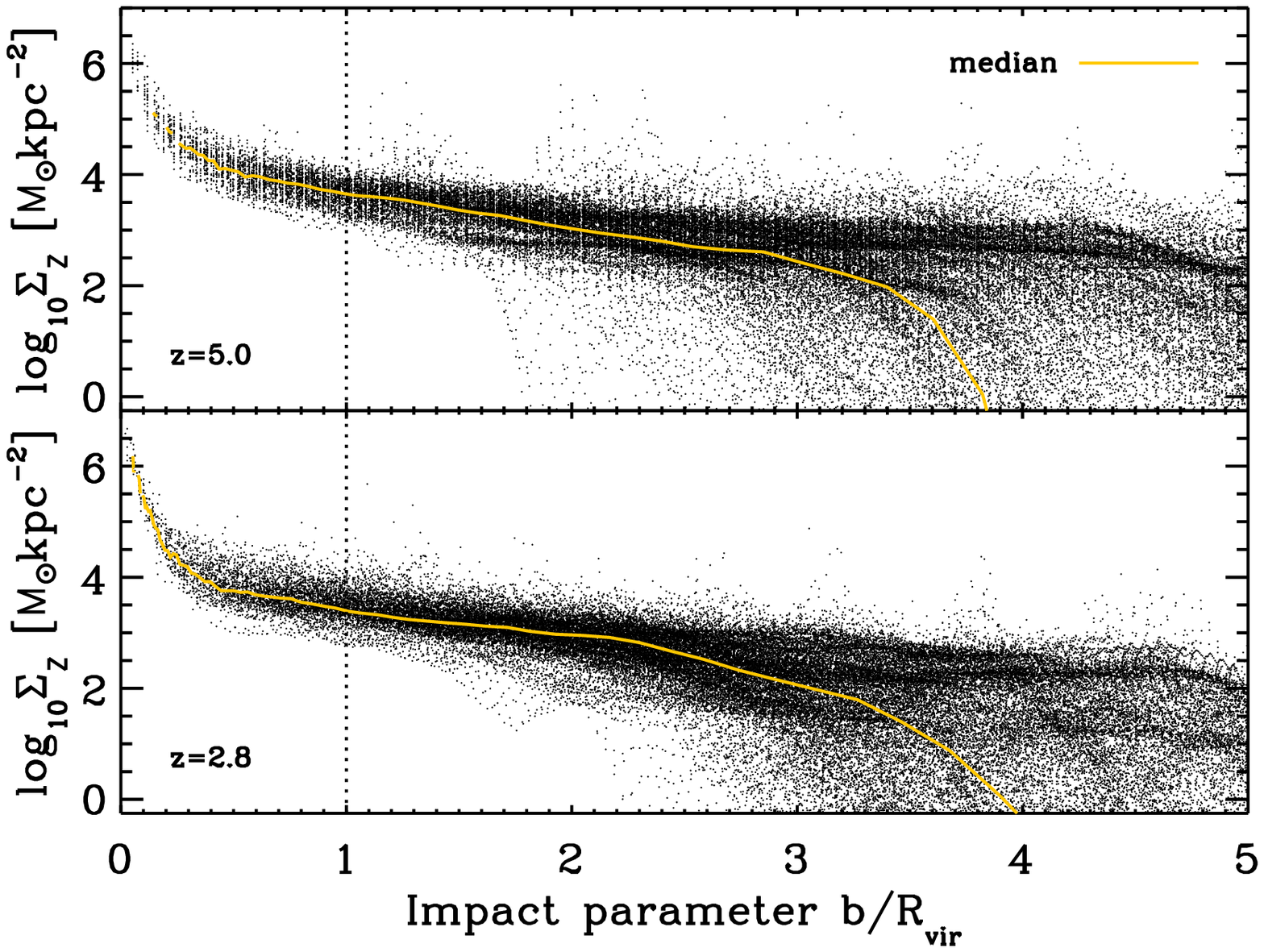}
\vspace{+0.3cm}
\caption{Total projected metal mass density, $\Sigma_Z$, along each sightline through Eris2's CGM, as a function of galactocentric impact parameter, $b$, in 
units of the virial radius. The metal mass density is plotted at two different redshifts, $z=2.8$ ($R_{\rm vir}=50$ kpc) and $z=5$ ($R_{\rm vir}=19$ kpc). 
The density of black points (one for each sightline) is a measurement of the covering factor of heavy elements in the simulation. The solid yellow line shows the 
median value of $\Sigma_Z$ at a given $b/R_{\rm vir}$. At the two redshifts, the metal bubbles around Eris2 appear scaled versions of each other.
}
\vspace{+0.3cm}
\label{fig8}
\end{figure}

Figures \ref{fig3a} and \ref{fig3b} show the total projected column densities in \HI, \CII, \CIV, \SiII, \SiIV, and \OVI\ for Eris2's $z=2.8$ CGM, 
and the columns of high ionization species that are cool and warm-hot, respectively. Figures \ref{fig4a} and \ref{fig4b} show the same columns 
for inflowing and outflowing CGM material only (gas particles are separated in inflowing and outflowing according to the sign of their radial 
velocities relative to the center of the main host), while the distribution of all enriched gas in the temperature-density plane is depicted in Figure \ref{fig5}.
  
Metal-enriched outflows are bipolar and perpendicular to the plane of the disk.
Heavy elements are clearly spread over a large range of phases, from cold star-forming material at $T<10^4$ K and $n \gta n_{\rm SF}=20$
atoms cm$^{-3}$ (corresponding to $\delta\equiv \rho/\rho_{\rm mean}\gta 10^{6.15}$ at $z=2.8$) to hot $T>10^6$ K low density $\delta\simeq 1$ intergalactic gas
that cannot cool radiatively over a Hubble time. \OVI\ is diffuse, has a large covering factor, and extends beyond $4R_{\rm vir}=200$ kpc from the center of Eris2.
\CIV\ absorption is much clumpier and less extended than \OVI, while low-ionization species like \CII\ and \SiII\ trace narrow inflowing streams as well
as dense outflowing clumps.  The covering fraction of low-ionization species declines faster than that of \OVI\ or \CIV.

Inflows and outflows coexist in Eris2, with about one third of all the gas within $R_{\rm vir}$ found to be outflowing. At $R_{\rm vir}$, 
the mean metallicity of inflowing gas is $0.05\,Z_\odot$, while the mean metallicity of outflowing gas is $0.56\,Z_\odot$ (the average is 
taken over all gas particles within a thin shell of radius $R_{\rm vir}$ and thickness 0.02 $R_{\rm vir}$). 
Inflowing enriched gas traces large-scale cold filaments that penetrate deep into the virial radius (see Fig. \ref{fig4a} and \S~\ref{coldflows}), as well 
as material raining back onto the disk as part of a ``galactic fountain".  
About 77\%, 32\%, 44\%, 66\%, 50\%, and 66\% of the total \HI, \OVI, \CIV, \CII, \SiIV, and \SiII\ mass within $2R_{\rm vir}$, respectively, is inflowing,  
High \HI\ column density absorption mainly traces the narrow cold inflowing streams. Outflowing material has smaller \HI\ columns and contributes 
to the covering factor of gas with $N_{\rm HI}> 10^{15.5}\cmm$.

The distribution of high-ionization species in the temperature-density plane is clearly bimodal.
At high densities and temperatures, ions are in collisional ionization equilibrium (CIE) and their fractional abundances are only a function of temperature. At lower densities, 
when photoionization is dominant, the equilibrium abundances become a function of density as well as temperature, and metals are typically in a higher ionization state 
at a given temperature than in CIE. \CII\ and \SiII\ ions trace primarily 
cool, $10^4 \lta T \lta 10^5$ K photoionized CGM gas at overdensities $\log \delta \simeq $2 (as well as dense, cold disk neutral material), which
extends as far as $2R_{\rm vir}$ but with a covering factor that decreases sharply beyond $R_{\rm vir}$ (Fig. \ref{fig3a}). \SiIV\ and \CIV\ are 
abundant in low density $\log \delta \gta 1$ photoionized gas. 
\CIV\ ions maintain a significant covering factor beyond $2R_{\rm vir}$ and can also be detected in warm-hot and denser $\log \delta\sim 2$ CGM gas within $R_{\rm vir}$.

Within 100 kpc, \OVI\ is largely collisionally ionized 
and traces CGM gas with $\log \delta\gta 1$ and $T>10^5$ K (in collisional ionization equilibrium, \OVI\ reaches its peak abundance fraction, $\sim 0.2$, at $3\times 10^5\,$K).
A cool, $T\lta 5\times 10^4$ K clumpy \OVI\ component with smaller covering factor becomes dominant at large galactocentric distances. This is relatively high column, 
moderate overdensity $\log \delta \sim 0.5-1.5$ gas that was enriched and expelled by nearby dwarfs and is photoionized by the cosmic UV background. 
The mass ratios of the hot versus cold \OVI\ populations are 17:1, 1.6:1, and 0.7:1 at distances $d<R_{\rm vir}$, $R_{\rm vir}<d<2R_{\rm vir}$, and $2R_{\rm vir}<d<3R_{\rm vir}$, 
respectively.

The covering factor of \OVI\ and \CIV\ absorbing material through Eris2's CGM is plotted in Figure \ref{fig6} as a function of column density.
The color coding marks the covering factor in different ranges of galactocentric impact parameter, while different symbols are used to indicate the 
median temperature of a given column of absorbing gas over all sightlines. The figure shows how \CIV\ absorbers with large covering factors are typically
cool (photoionized), while the opposite is true for \OVI. A hot \CIV\ component is also present at small impact parameters and low columns. 

\begin{figure*}
\centering
\includegraphics[trim=0.3in 4.7in 0in 0.8in,clip,width=1.1\textwidth]{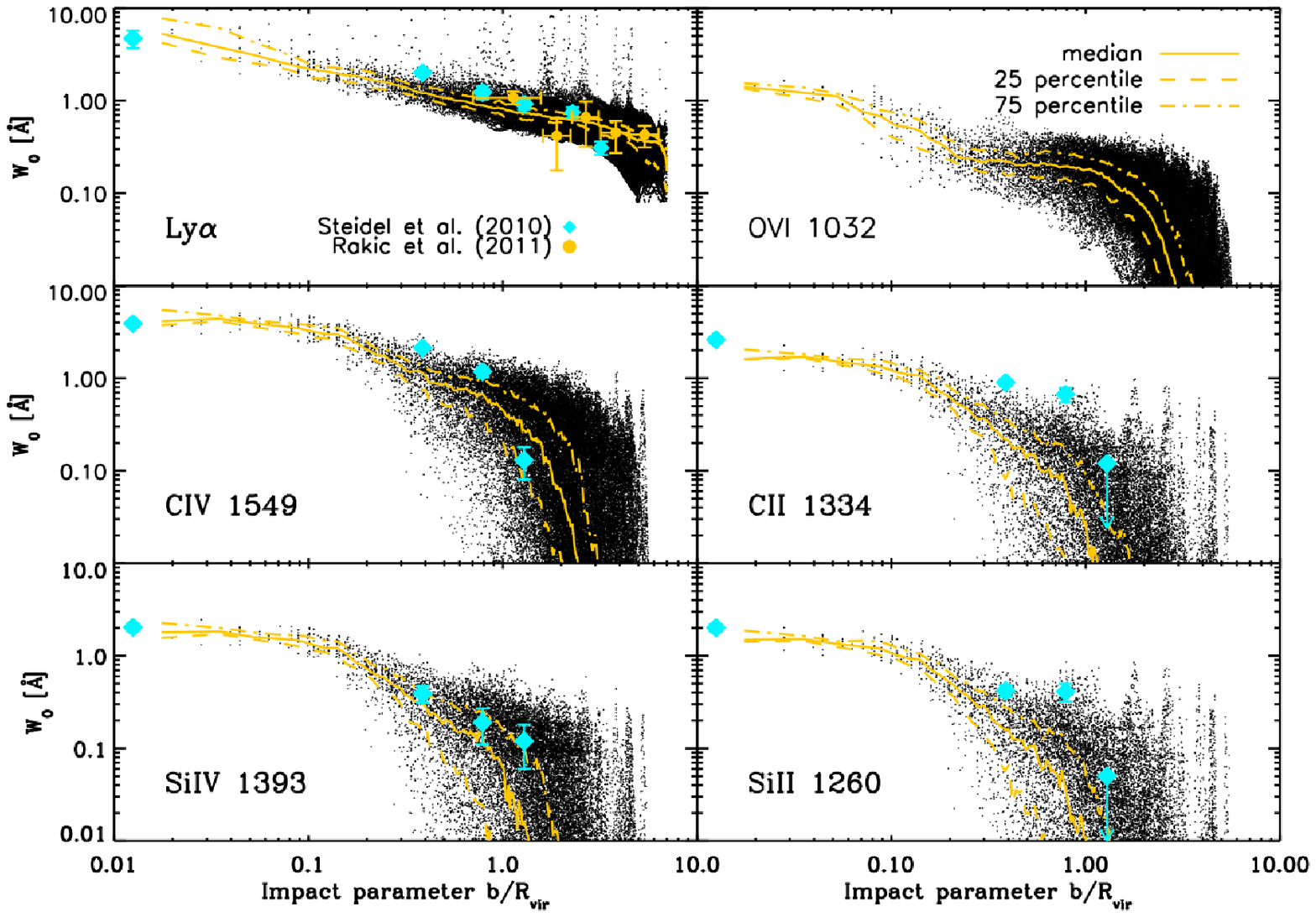}
\vspace{0.0cm}
\caption{Interstellar absorption line strengths of the 1216 \Lya, 1334 \CII, 1549 \CIV, 1260 \SiII, 1393 \SiIV, and 1032 \OVI\ 
transitions as a function of galactocentric impact parameter $b$ in physical kpc. Values for \CIV\ include both components of the 
$\lambda\lambda$1548,1550 doublet. The cyan points at $b$=1, 31, 63, and 103 kpc are data from \citet{Steidel10}, with downward arrows 
on points indicating upper limits. The yellow points in the top left panel are data from \citet{Rakic12}. 
The impact parameter has been normalized to $R_{\rm vir}=80$ kpc (see text for details). 
The black dots show the rest-frame equivalent widths measured along artificial sightlines through the CGM of our Eris2 simulation at $z=2.8$, and the yellow lines
marks the median and the 25th and 75th percentiles. 
}
\label{fig9}
\end{figure*}

The formalism developed in this Section can be finally used to compute the optical depth of the hydrogen \Lya\ $\lambda$1216, \CII\ $\lambda$1334, 
\CIV\ $\lambda\lambda$1548,1550, \SiII\ $\lambda$1260, \SiIV\ $\lambda$1393 and \OVI\ $\lambda$1032 transitions. 
Figures \ref{fig7a} and \ref{fig7b} show artificial absorption spectra along two lines of sight through the CGM at impact parameters $b=23$ kpc and $b=34$ kpc 
from Eris2's center, plotted on a galactocentric velocity scale. Column densities range from $8.7\times 10^{12}\,\cmm$ for \SiII\ to $1.3\times 10^{17}\,\cmm$ for hydrogen \Lya, 
and individual components span a radial velocity range from about $-200$ to $+200\,\kms$. Velocity profiles are complex, with ouflowing and inflowing 
gas detected at positive (redshifted) and negative (blueshifted) projected velocities from the near and the far side of the galaxy. The simultaneous presence, 
in velocity space, of features from singly ionized species as \CII\ and \SiII\ along with higher-ionization species reflects the multi-phase structure of the simulated CGM. 
The sightline depicted in Figure \ref{fig7b} goes through cool, metal-enriched gas infalling on the near side of the galaxy. The inflowing component is distributed 
in physical space between $R_{\rm vir}$ and $2 R_{\rm vir}$, has overdensities $\log \delta \gta 2$, metallicities in excess of $0.03\,Z_\odot$, and is part of 
the metal-enriched zone around a dwarf satellite companion accreting along a filament. We will discuss the observability of cold acccretion streams via low-ionization metal lines
in the next section. The spectrum shows several outflowing components, including solar-metallicity photoionized gas at $v\sim -150\,\kms$ that is being ejected by the main host, as well as 
pristine IGM gas at $v\sim 0\,\kms$ that is located $\gta 240$ kpc from the galaxy center. 

A detailed analysis of interstellar absorption line strengths as a function of redshift will be the subject of another paper. 
Here, it is important to keep in mind that, when properly scaled, the property of Eris'2 CGM are not changing rapidly with cosmic time. Figure \ref{fig8} shows
the total projected metal mass density, $\Sigma_Z$, along each sightline, as a function of galactocentric impact parameter in units of the virial radius, 
$b/R_{\rm vir}$. The metal mass density is plotted at two different redshifts, $z=2.8$ ($R_{\rm vir}=50$ kpc) and $z=5$ ($R_{\rm vir}=19$ kpc). 
The density of black points (one for each sightline) is a measurement of the covering factor of metal-enriched material in the simulation.
At the two redshifts, the metal bubbles around Eris2 appear scaled versions of each other.

\section{Comparison with the observations}\label{comparison}

\subsection{Circumgalactic metals at $z$=2-3}

The close pair sample of galaxies at $\langle z\rangle=2.2$ of \citet{Steidel10}, drawn from a spectroscopic survey of ``BX" UV-color 
selected objects, provides a robust map of cool circumgalactic gas at galactocentric impact parameters $b=3-125$ kpc. Objects in this sample 
have been well-characterized in terms of their spatial clustering and stellar population parameters \citep{Adelberger05,Erb06b}.
A good match to both the observed clustering strength and the space density of the ``BX" galaxies to dark matter halos in the Millenium simulation
is obtained for an average halo mass of $\langle M_{\rm vir}\rangle=9 \times 10^{11}\,\msun$ \citep{Conroy08}. The virial radius for a halo at the average mass 
that formed several $10^8$ yr prior to the epoch of observation is $R_{\rm vir}=80$ kpc (Steidel \etal 2010),
close to the scale at which low-ionization absorption lines become weaker. 
This suggests that beyond $R_{\rm vir}$ absorbers may become too diffuse and highly ionized to produce significant columns of low-ionization metals
\citep{Steidel10}. With a virial mass at $z=2.8$ of $M_{\rm vir} = 2.6 \times 10^{11}\,\msun$ and a radius $R_{\rm vir} = 50$ kpc, Eris2 is a smaller counterpart to 
\citet{Steidel10} close pair sample galaxies.\footnote{This is also true in the luminous component, as the mean stellar mass of the spectroscopic 
sample, $\langle M_*\rangle=(3.6\pm 0.4)\times 10^{10}\,\msun$ \citep{Erb06b}, is 2.5 times larger than in Eris2.}~For a proper comparison between simulation 
data and observations, we have therefore scaled the impact 
parameter with the virial radius of the mean halo in the pair sample (80 kpc) and with the Eris2's virial radius in the simulation, and plotted the 
observed and predicted line strengths against the correspondingly normalized impact parameters. Along every simulated sightline, the rest-frame equivalent
widths $W_0$ of \CII\ 1334, \CIV\ 1549 (including both components of the doublet), \SiII\ 1260, and \SiIV\ 1393 are shown in Figure \ref{fig9}
versus $b/R_{\rm vir}$, together with the values observed in the composite spectra of foreground galaxies (placed at $\langle b\rangle=1$ kpc)   
and background galaxies (at $\langle b\rangle=31,63,103$ kpc) by \citet{Steidel10}. The density of black points is a measurement of the covering factor of 
absorbing material at a given impact parameter in the simulation. We plot $W_0$ along every artificial line of sight 
rather than an average over all sightlines in order to underline the dispersion in the measured line strengths as well as the contribution
of satellite galaxies (the narrow ``spikes" in equivalent width seen at $b/R_{\rm vir}>1$). For each species, the strength of absorption declines rapidly  
around $b=$1-2 $R_{\rm vir}$. The agreement between the simulation and the observations is rather good: except perhaps for a slight deficit in the 
covering factor of \CII\ and \SiII\ around $\sim 0.8 R_{\rm vir}$ and an excess of \CIV\ at large impact parameters , Eris2's simulated CGM broadly reproduces the 
line strengths-impact parameter trends observed by \citet{Steidel10}. In particular, the covering factor of absorbing 
material declines less rapidly with impact parameter for \Lya\ and \CIV\ (and \OVI) compared to \CII, \SiIV, and \SiII, with \Lya\ remaining 
strong ($W_{\rm Ly\alpha}>300\,$m\AA) 
to $\gta 5R_{\rm vir}=250$ kpc. At small impact parameters, where lines are mostly saturated, the equivalent width along a sightline is modulated by the 
velocity structure of the absorbing material. The kinematics of Eris2's inner CGM appears therefore to be consistent with the observations.
A comparison between Figure \ref{fig9} and Figure 15 of \citet{Fumagalli11} shows that the predicted interstellar \SiII\ equivalent widths are about an order of magnitude 
larger in Eris2 than in \citet{Fumagalli11}, and therefore that strong galactic outflows are an essential ingredient 
for a quantitative modeling of the CGM.

\begin{figure}
\centering
\includegraphics[width=0.499\textwidth]{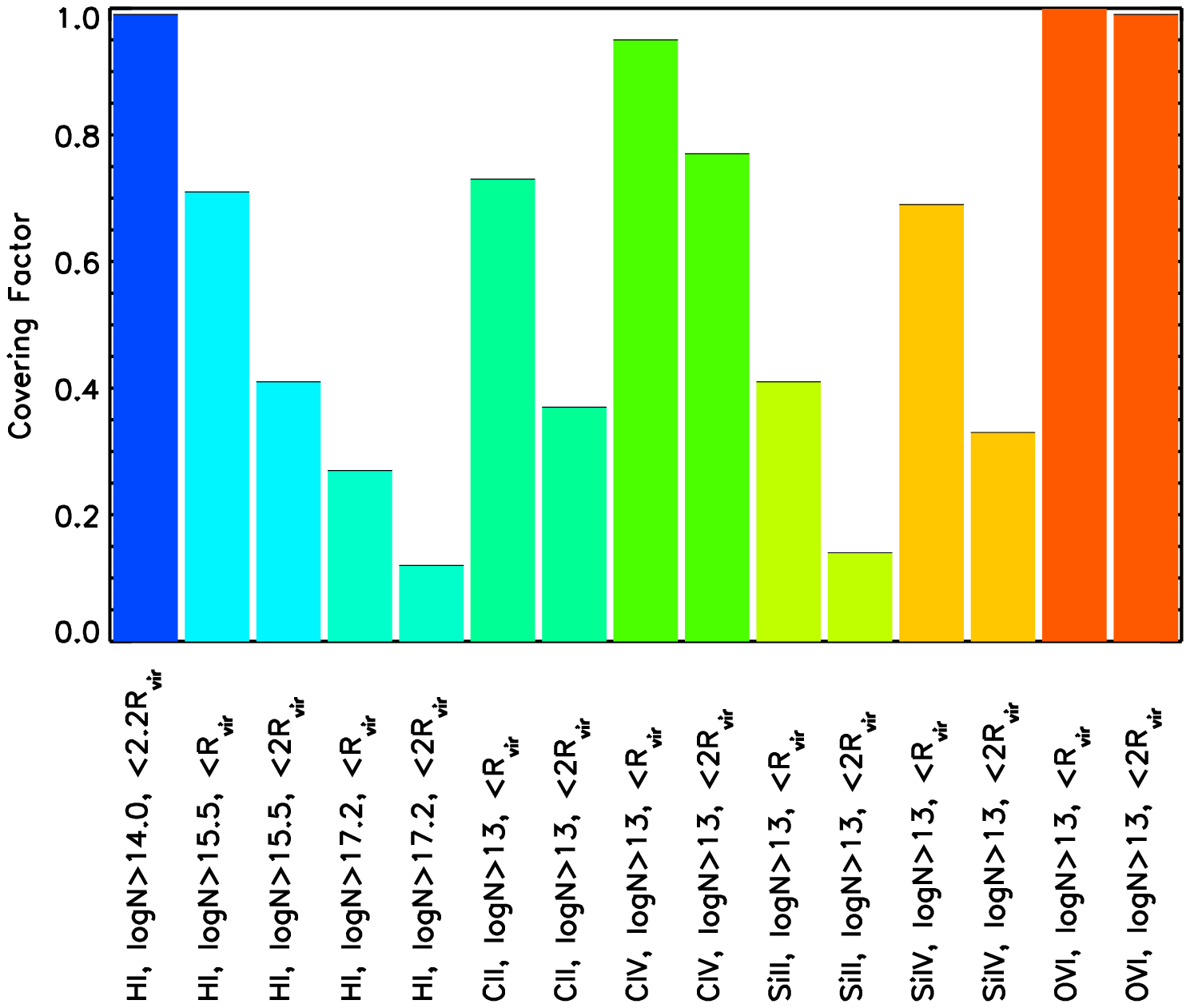}
\vspace{+2.0cm}
\caption{
The covering factor at different galactocentric impact parameters of absorbers of various species and column densities (in cm$^{-2}$) in Eris2's CGM.
%{\it Right panel:} The gas mass, total metal mass, and mass of individual species that is inflowing and outflowing, as well as cold and hot.
} 
\label{fig10}
\vspace{+0.0cm}
\end{figure}

\subsection{Circumgalactic \iHI\ and cold flows}
\label{coldflows}

Figure \ref{fig9} also shows the equivalent width of hydrogen \Lya\ absorption as a function of impact parameter for the galaxy-galaxy 
pair sample of \citet{Steidel10} as well as the QSO-galaxy pair sample of \citet{Rakic12}. Again, the simulation agrees well with the observations, as 
\Lya\ remains strong ($W_{\rm Ly\alpha}>300\,$m\AA) to $5R_{\rm vir}=250$ kpc. Infall of cool gas via narrow streams, directly onto the central regions of galaxies, 
is the mode of gas accretion that is predicted to feed star formation in high redshift galaxies like Eris2 \citep[e.g.][]{Keres05,Dekel06,Ocvirk08,Agertz09,Dekel09}.
Observational signatures of cold accretion streams at high redshift are key for testing our understanding of the flows of matter, energy, and metals into and out of galaxies. 
Absorption by neutral hydrogen (and, as we shall argue below, by low-ionization metal lines) is a promising way to observe the cold streams.
Recent observations of $2.0\lta z\lta 2.8$ star-forming galaxies near a QSO sightline (the Keck Baryonic Structure Survey) show that
the covering fraction of absorbers of various $N_{\rm HI}$ columns around galaxies ranges from $90\pm 9$\% ($<R_{\rm vir}$) to $68\pm 9$\%   
($<2 R_{\rm vir}$) for $\log N_{\rm HI}>15.5$, and from $30\pm 14$\% ($<R_{\rm vir}$) to $28\pm 9$\% ($<2 R_{\rm vir}$) for 
Lyman Limit Systems (LLSs) with $\log N_{\rm HI}>{17.2}$ \citep{Rudie12}. 

\begin{subfigures}

\begin{figure*}
\centering
\includegraphics[width=0.95\textwidth]{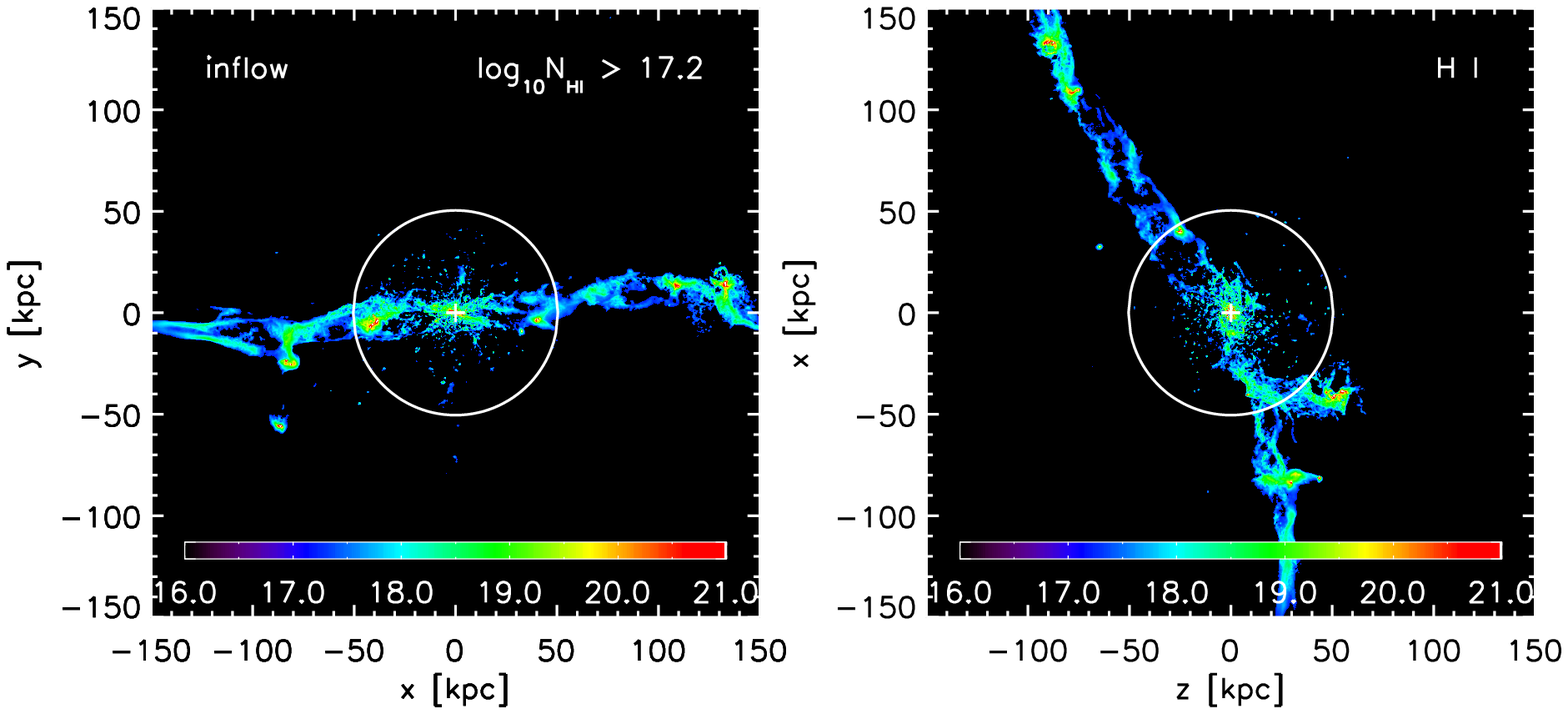}
\vspace{+0.2cm}
\caption{Cold accretion streams in Eris2: a map of hydrogen column density showing inflowing, optically thick material with $\log N_{\rm HI}>{17.2}$ for
2 different projections. Eris2's virial radius is marked by the white circle. 
} 
\label{fig11a}
\vspace{+0.0cm}
\end{figure*}

\begin{figure*}
\centering
\includegraphics[width=0.95\textwidth]{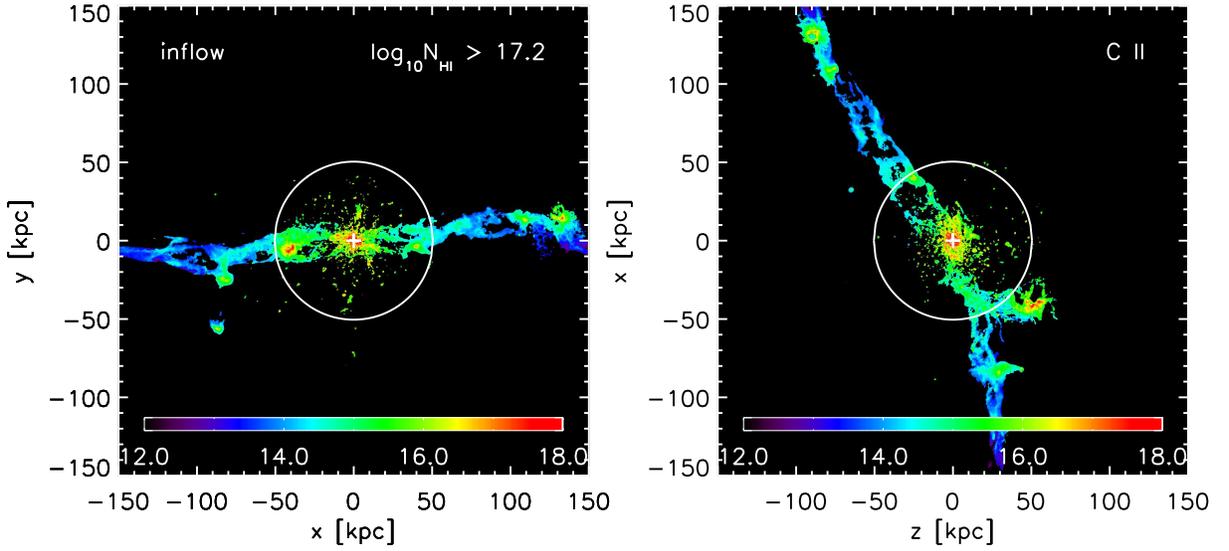}
\vspace{+0.2cm}
\caption{Same as Fig. \ref{fig11a}, showing the \CII\ column for optically thick inflowing gas. 
} 
\label{fig11b}
\vspace{+0.0cm}
\end{figure*}

\end{subfigures}

The simulation (see Fig. \ref{fig10}) yields covering factors of 71\% and 27\% ($<R_{\rm vir}$) and of 41\% and 12\% ($<2 R_{\rm vir}$) 
for $\log N_{\rm HI}>15.5$ and $>17.2$, again in reasonable agreement with the data, albeit perhaps on the low side for the largest impact parameters. 
According to \citet{Rudie12}, the incidence of absorbers with $\log N_{\rm HI}>14.5$ is found to be higher near galaxies than in the general IGM.
The covering factor of $\log N_{\rm HI}>14$ systems within 2.5 $R_{\rm vir}$ is observed to be $92\pm 5$\% vs. 99\% in Eris2. Eris2's covering factor of LLSs appears 
comparable or higher than those predicted at high redshift by \citet{Faucher11} and \citet{Fumagalli11} for galaxies in the same mass range. 
Both these groups ran high-resolution cosmological hydrodynamical simulations that do not generate strong galactic ouflows (\citealt{Fumagalli11} using 
the AMR technique, \citealt{Faucher11} using SPH as in Eris2), combined with radiative transfer. Because our hydrodynamical simulations lack proper 
ionizing radiative transfer, we cannot correctly address the covering factor of the self-shielded, thicker Damped \Lya\ (DLA) absorbers with $\log N_{\rm HI}\gta 20.3$. 

\begin{figure}
\centering
\includegraphics[width=0.49\textwidth]{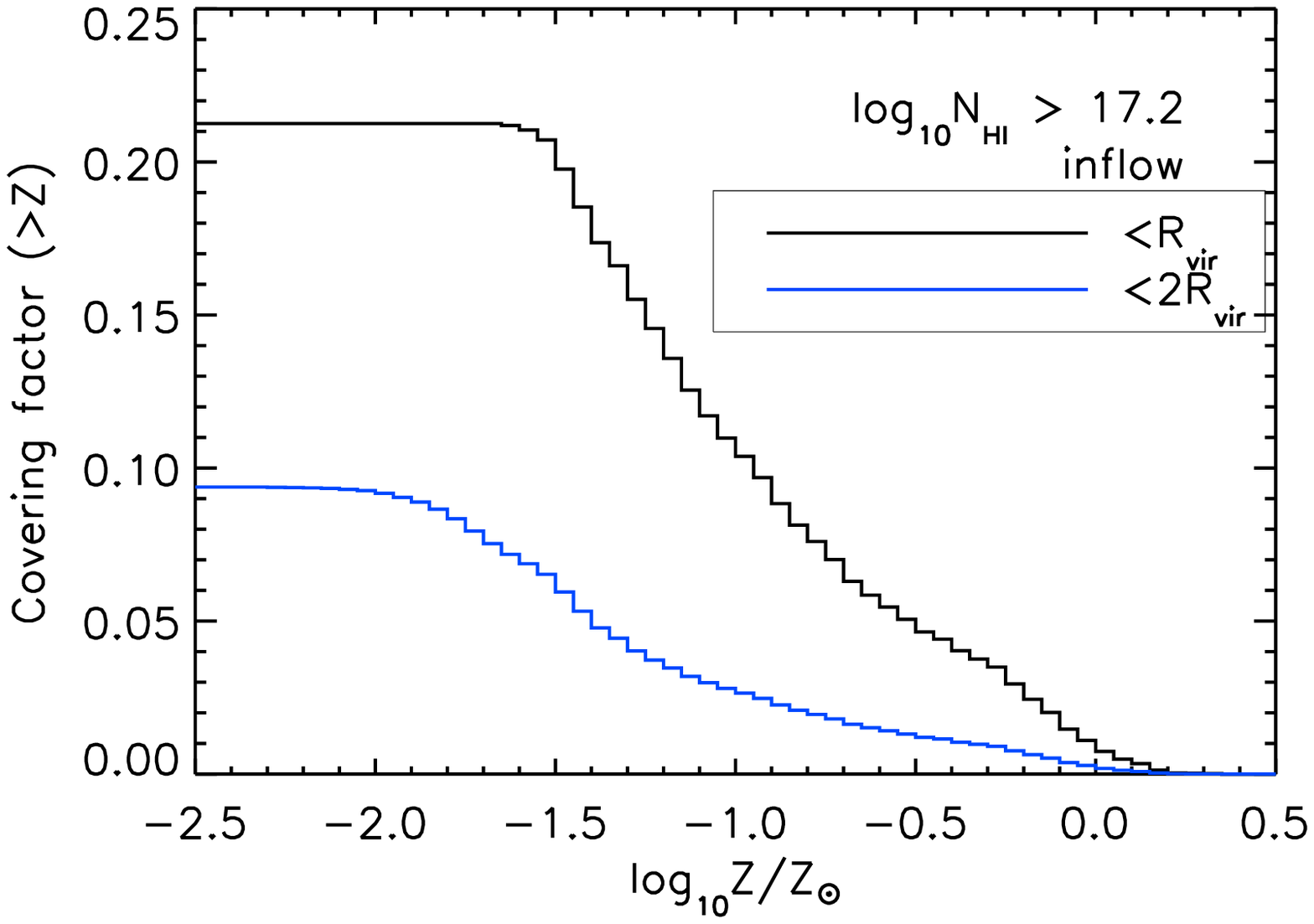}
\vspace{+0.2cm}
\caption{Cold, optically thick streams in Eris2 are metal-enriched. The cumulative covering factor of $\log N_{\rm HI}>17.2$ inflowing material above a metallicity $Z$ 
is plotted within $R_{\rm vir}$ and $2 R_{\rm vir}$. Streams within $R_{\rm vir}$ ($2 R_{\rm vir}$) are enriched above 0.03 (0.01) solar by 
previous episodes of star formation in the main host and in dwarf satellite galaxies. 
} 
\label{fig12}
\vspace{+0.0cm}
\end{figure}

It is interesting at this stage to look at the properties of cold streams. About 90\% of all optically thick absorbing gas within 1-2 virial radii is 
found to be inflowing with radial velocities $\lta 150-200\,\kms$. Figures \ref{fig11a} and \ref{fig11b} shows two projections of all {\it inflowing} material 
with $\log N_{\rm HI}> 17.2$, as traced by \HI\ and \CII, respectively. The streams can be clearly seen penetrating deep inside the virial radius and delivering 
cold gas to the central galaxy. Enriched to metallicities above 0.01 solar by previous episodes of star formation in the main host and nearby dwarfs \citep{Shen12,Fumagalli11},  
the streams give origin to strong ($N_{\rm CII}>10^{13}\cmm$) \CII\ absorption with a covering factor of 22\% within $R_{\rm vir}$ and 10\% within $2 R_{\rm vir}$, which
should make the presence of cold flows detectable with metal absorption lines. We find no substantial 
suppression of the cold accretion mode caused by galactic-scale mass outflows. The mass inflow rate of cold ($T<10^5$ K) gas into Eris2 is $\dot M_{\rm cold}=18\,\mdot$  
(measured at $R_{\rm vir}$), in agreement with the results of \citet{Keres09} for galaxies in the same redshift and mass range as Eris2, and comparable to Eris2's star
formation rate. About 35\% of the cold gas mass is brought in by dwarf satellite galaxies. The cold mode dominates over the hot mode, which is characterized by an accretion 
rate of $\dot M_{\rm hot}=5\,\mdot$. The distribution of metallicity, column density weighted, in the entire sample of inflowing LLSs within $2R_{\rm vir}$ is centered around a 
median value of $0.07\,Z_\odot$, significantly higher than found in the simulations of \citet{Fumagalli11,Kimm11}. Nearly all inflowing LLSs have metallicities above 0.01 solar (Figure \ref{fig12}).

\begin{figure*}
\centering
\includegraphics[width=0.95\textwidth]{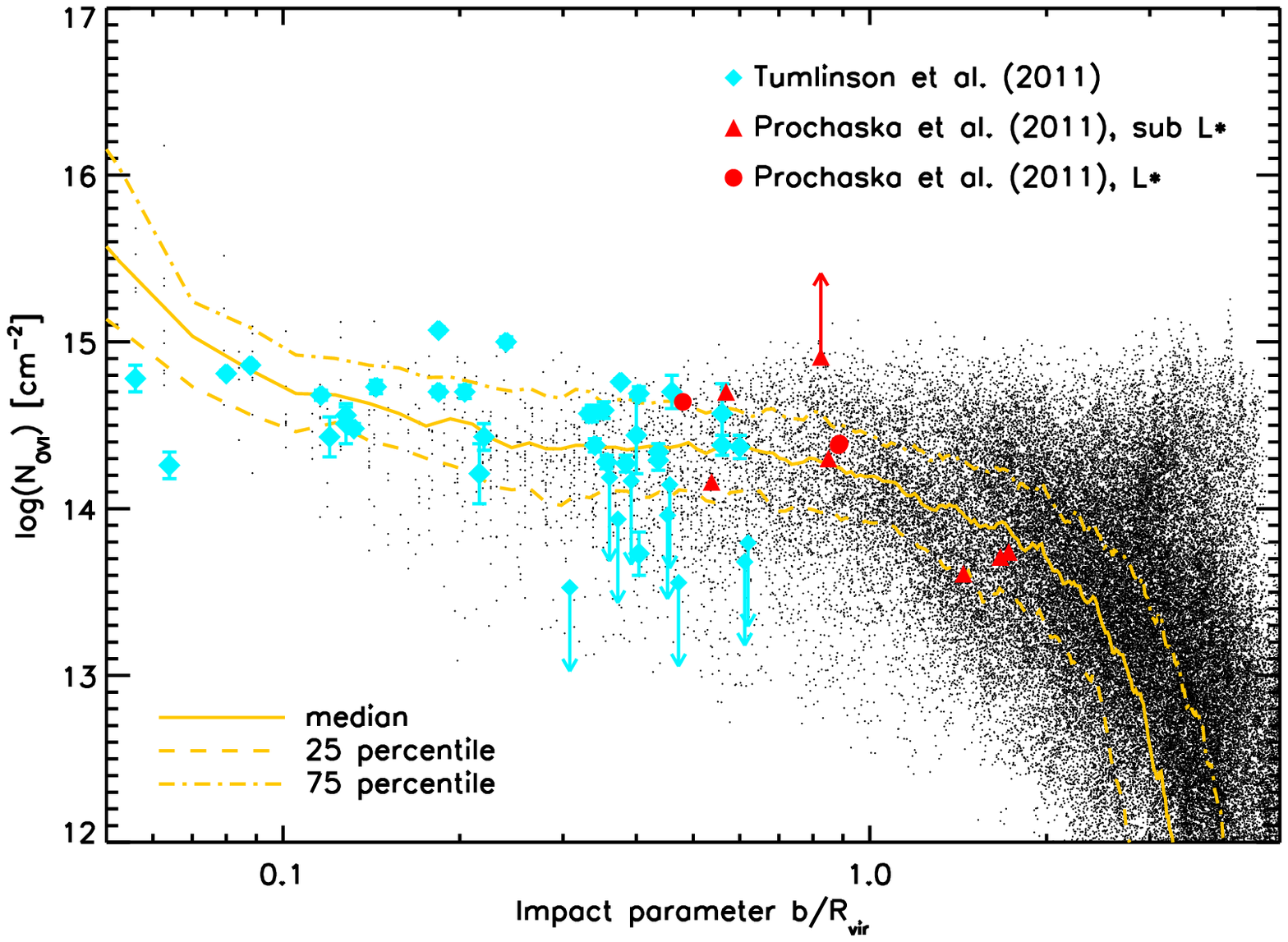}
\vspace{0.0cm}
\caption{The \OVI\ column density vs. impact parameter for the star-forming subsample of \citet{Tumlinson11} ({\it cyan squares}) 
and the $L_*$ ({\it red dots}) and sub-$L_*$ ({\it red triangles}) subsamples of \citet{Prochaska11} of low-redshift galaxies. 
The $L_*$ data of \citet{Prochaska11} and those of \citet{Tumlinson11} have been scaled assuming a virial radius of 250 kpc. The sub-$L_*$ data 
of \citet{Prochaska11} have been scaled assuming a virial radius of 160 kpc. The black dots show the \OVI\ column density measured along artificial sightlines 
through the CGM of our Eris2 simulation at $z=2.8$, and the yellow lines marks the median and the 25th and 75th percentiles. 
} 
\label{fig13}
\vspace{+0.0cm}
\end{figure*}

\subsection{The \OVI\ halos of star-forming galaxies}

As already discussed in the previous section, as a consequence of the SN blastwave-driven galactic outflows generated in our simulations,
the central galaxy is surrounded by a large halo of collisionally ionized \OVI. Ubiquitous, large (150 kpc) halos of \OVI\ around 
low-redshift star-forming galaxies have been recently detected in a survey with the Cosmic Origins Spectrograph onboard of the {\it HST} \citep{Tumlinson11}, 
and it is interesting to compare their properties with our simulation. The \citet{Tumlinson11} sample 
includes galaxies with stellar masses $9.5<\log\,(M_*/\msun)<11.2$ and specific star formation rates $-11<\log\,({\rm sSFR/yr^{-1}})<-9$.  
Eris2's stellar mass and sSFR at $z=2.8$ are in the middle and on the high side of such ranges, respectively. We have computed the 
total \OVI\ column along simulated sightlines at different impact parameters and plotted the results in Figure \ref{fig13}  
together with the \citet{Tumlinson11} and \citet{Prochaska11} low-redshift observations. Our high-redshift simulated galaxy again matches the trends 
observed in similar systems at low redshfts, with typical column densities of $N_{\rm OVI}\gta 10^{14}\,\cmm$ and a near unity covering factor are 
maintained all the way out to 100-150 kpc. 

\section{Summary} 

Recent observations of the CGM of star-forming galaxies at high and low redshift have called for more stringent comparisons with hydrodynamical simulations of galaxy 
formation. The high-resolution ``zoom-in" simulation presented here, Eris2, adopts a blastwave scheme for supernova feedback which, in combination with a high gas 
density threshold for star formation, has been shown to generate large-scale galactic outflows and be key in producing realistic dwarf galaxies and late-type 
massive spirals \citep[e.g.][]{Governato10,Guedes11}, in turning dark matter cusps into cores \citep[e.g.][]{Pontzen12}, and in enriching the CGM of high redshift LBGs 
\citep{Shen12}. Simulations of star-forming galaxies run to $z=0$ have recently stressed the importance of strong stellar feedback in reproducing the \OVI\ hot gaseous 
halos observed in local galaxies \citep{Stinson11}. Eris2 includes metal-dependent radiative cooling, a model for the diffusion of metals and thermal energy, and a local UV radiation 
field added in post-processing, and is ideally suited for a detailed study of the kinematics, thermal and ionization state, and spatial distribution of metal-enriched gas in the CGM
of massive galaxies at redshift $\sim 3$. The results presented in this paper do not depend sensitively on metal diffusion: a twin simulation without diffusion
produces a slightly clumpier CGM, but the impact on the covering factor of metal ions is small.    

We have generated synthetic spectra by drawing sightlines through the simulated CGM at different galactocentric impact parameters, 
and compared the theoretical interstellar absorption line strengths with the observations. We emphasize again that the parameters of the simulation 
have not been tuned to provide a fit to observations of the CGM. In our examination we have found that:

\begin{itemize}

\item Eris2's CGM at $z=2.8$ contains multiple phases having a wide range of physical conditions, with half of all gas-phase metals locked in a warm-hot 
component at $T>10^5\,$K. Outflows and inflows coexist, and about one third of all the gas within $R_{\rm vir}$ is outflowing. Inflows bring in material 
along optically thick ``cold" streams. The streams are enriched to metallicities above 0.01 solar by previous episodes of star formation in the main host 
and in nearby dwarfs. 

\item The \CII\ and \SiII\ ions trace primarily cool, $10^4<T<10^5$ K photoionized CGM gas at overdensities $\log \delta \simeq $2 extending as far as $2R_{\rm vir}$,
but with a covering factor that decreases sharply beyond $R_{\rm vir}$. The distributions of high-ionization species are clearly bimodal, with ions in CIE at high 
densities and temperatures, and with photoionization becoming dominant at low densities. \CIV\ ions maintain a significant covering factor beyond $2R_{\rm vir}$ and can 
also be detected in warm-hot and denser $\log \delta\sim 2$ CGM gas within $R_{\rm vir}$. The \OVI\ diffuse halo extends beyond $4R_{\rm vir}=200$ kpc from the center. 
Within 100 kpc, \OVI\ is largely collisionally ionized and traces CGM gas with $\log \delta\gta 1$ and $T>10^5$ K. A cool, clumpy \OVI\ component, photoionized by the
cosmic UV background, becomes increasingly important at larger distances. 

\item Synthetic spectra generated by drawing sightlines through the simulated CGM produce interstellar absorption line strengths of \Lya, \CII, \CIV, \SiII, and \SiIV\ 
as a function of galactocentric impact parameter (scaled to the virial radius) that are in good agreement with those observed at high-redshift by Steidel \etal 
(2010) (and \citealt{Rakic12} in the case
of \Lya). The covering factor of absorbing material is found to decline less rapidly with impact parameter for \Lya\ and \CIV\ (and \OVI) compared to \CII, \SiIV, and 
\SiII. \Lya\ remains strong to $\gta 5R_{\rm vir}=250$ kpc. At small impact parameters, where lines are mostly saturated, the equivalent width along a sightline 
is modulated by the velocity structure of the absorbing material. The kinematics of Eris2's inner CGM appears therefore to be consistent with the observations.
Note that, contrary to the spherically outflowing wind model of \citet{Steidel10}, in our simulations inflowing material contributes significantly to the absorption
in the $W_0-b$ plane.   

\item At small impact parameters, the local galactic UV flux increases the abundance of high-ionization species like \CIV\ and \SiIV, and decreases that of \CII, \SiII, and \HI. 
\OVI\ is largely collisionally ionized within 100 kpc, so the effect of the local UV on \OVI\ is small.  More quantitatively, within $R_{\rm vir}$, galactic radiation increases the total amount of \OVI\ by 1\%, \CIV\ by 128\%, and \SiIV\ by 146\%, while it decreases \CII\ by 53\%, \SiII\ by 51\%, and \HI\ by 87\%. 
Nevertheless, the effect on the equivalent widths of \Lya\ and metal transitions is small. This is because absorption lines are saturated in the local CGM and their 
equivalent widths are mainly determined by the kinematics and covering factor of the gas rather than by its column density. 

\item
The fraction of sightlines within one virial radius that intercept optically thick, $N_{\rm HI}>10^{17.2}\,\cmm$ gas is 27\%, in good agreement 
with recent observations by \citet{Rudie12} who find $30\pm 14$\%. Within $2R_{\rm vir}$, our covering factor is 12\%, below the $28\pm 9$\% observed by
\citet{Rudie12}. The corresponding covering factors for $\log N_{\rm HI}>15.5$ are $(71\%,41\%)$ versus $(90\pm 9\%,68\pm 9\%)$ found in the observations.

\item Optically thick material is found to trace cold accretion streams that penetrate deep inside the virial radius and deliver cold gas to the central 
galaxy. The streams have metallicities above 0.01 solar. \CII\ absorption with $N_{\rm CII}>10^{13}\cmm$ arises from such streams with a covering factor of 
22\% and 10\% within one and two virial radii, respectively, making the presence of cold flows detectable with metal absorption lines. There is no substantial 
suppression of the cold accretion mode caused by galactic outflows.

\item The \OVI\ halo gives origin to CGM absorption with a typical column density of $N_{\rm OVI}\gta 10^{14}\,\cmm$ and a near unity covering factor that 
is maintained all the way out to 150 kpc. This matches the trends recently observed in star-forming galaxies at low redshift by \citet{Tumlinson11} and \citet{Prochaska11}.  
\end{itemize}

While our zoom-in simulations of this single system appear then to reproduce quantitatively the complex baryonic processes that determine the exchange of matter, energy, and metals 
between galaxies and their surroundings, we acknowledge a number of caveats to the work presented here. Radiation transport is not explicitly present in our simulations, and while 
we have checked the robustness of our conclusion with a self-shielding approximation used in post-processing, simulations that incorporate some form of radiative transfer are 
in the making. Eris2's interstellar absorption features do not reach the maximum velocities $|v_{\rm max}|\simeq 800\,\kms$ observed in LBGs by \citet{Steidel10}, and   
zoom-in simulations of more massive systems are needed in order to fully test our galactic outflow model against observations of the high-redshift CGM. 
The clumpiness of Eris2's CGM may, at least partly, be a numerical artifact of the SPH technique. A clumpy, 
two-phase medium could be produced by thermal instabilities in the warm/hot halo around the main host \citep{Maller04,Kaufmann06,Kaufmann09}, although whether or 
not the conditions for 
thermal instabilities to operate are met in real galaxies is still a controversial issue \citep{Binney09}. The process has been observed in idealized simulations in 
which gas cools out of an initially uniformly hot corona and the instability is seeded by Poisson-noise density fluctuations in the initial particle distributions. 
\citet{KeresH09} have argued that clouds arise naturally in realistic cosmological hydrodynamical simulations as a result of a combination of Rayleigh-Taylor instabilities 
at the interface between cold accretion flows and the hot phase and thermal instabilities. Moreover, hot outflows and winds from supernovae launched from the disk might 
induce strong, large wavelength fluctuations that are not easily damped, and may thus act as the seeds to thermal instability in the CGM. 
These seeding mechanisms, as well as perturbations by satellites, may all play a role in our simulations. A detailed analysis of the clumpiness of Eris2' CGM 
as the possible result of such instabilities will be the topic of future work.  

\acknowledgments
Support for this work was provided by the NSF through grant AST-0908910 and OIA-1124453, and by NASA through grant 
NNX12AF87G (P.M.). Resources supporting this work were provided by the NASA High-End Computing (HEC) Program through the 
NASA Advanced Supercomputing (NAS) Division at Ames Research Center. J.G. was partially funded by the ETH Zurich Postdoctoral 
Fellowship and the Marie Curie Actions for People COFUND Program. We acknowledge useful discussions on the topics of the paper 
with A. Aguirre, S. Bertone, A. Dekel, M. Fumagalli, and J. Werk.

\end{document}